\documentclass[a4paper,12pt]{article}

\usepackage[centertags]{amsmath}
\usepackage{amssymb}

\usepackage{graphicx}
\usepackage{epsfig}
\usepackage{ulem}
\usepackage[english]{babel}
\usepackage{array}
\usepackage{amsthm}
\usepackage{latexsym}

\usepackage{yfonts}
\usepackage{mathrsfs}
\usepackage[mathcal]{euscript}

\pdfoutput=1
\usepackage{epsfig}
 \usepackage{jheppubm}
\usepackage{mathdots}
\usepackage{MnSymbol}
\usepackage{multirow}

\definecolor{darkraspberry}{rgb}{0.53, 0.15, 0.34}

\definecolor{darkblue}{rgb}{0., 0, 1}

\definecolor{dgreen}{rgb}{0.,0.6,0.}

\newcommand{\nn}{\nonumber}
\newcommand{\be}{\begin{equation}}
\newcommand{\ee}{\end{equation}}
\newcommand{\bea}{\begin{eqnarray}}
\newcommand{\eea}{\end{eqnarray}}

\newcommand{\cD}{{\cal D}}

\newcommand{\cI}{{\cal I}}

\newcommand{\cL}{{\cal L}}

\newcommand{\cR}{{\cal R}}

\newcommand{\Um}{{\mathscr{U}}}
\newcommand{\Vm}{{\mathscr{V}}}
\newcommand{\Wm}{{\mathscr{W}}}

\title{A Note on Islands in Schwarzschild \\ Black Holes 
}
\author{Irina Aref'eva and Igor Volovich}

\affiliation{Steklov Mathematical Institute, Russian Academy of Sciences,\\Gubkina str. 8, 119991, Moscow, Russia}
\emailAdd{arefeva@mi-ras.ru}
\emailAdd{volovich@mi-ras.ru}

\abstract{We  consider evaporation of the Schwarzschild black hole and note that    island 
configurations do not provide a bounded entanglement entropy.
The same remark is valid also for some other static black holes.  A proposal  for improving the situation is discussed.\\

}
\begin{document}
\maketitle

\section{Introduction}

The entropy of Hawking radiation of black holes grows up to infinity during evaporation and it is a manifestation of the information paradox \cite{Haw1,Haw2}. 
This increase contrasts with Page's hypothetical behavior, in which entropy decreases after the so-called  Page time \cite {Page:1993wv,Page:1976df, Page:2013dx} and which ensures the unitarity of quantum mechanics.\\

An approach to treating the  black hole information problem  was suggested in \cite{Penington:2019npb,
Almheiri:2019psf,
Almheiri:2019hni,
Almheiri:2019yqk,
Almheiri:2020cfm}, where the ``island formula''
for the entanglement entropy of Hawking radiation based on quantum extremal surfaces \cite{Engelhardt:2014gca} has been proposed and it has been argued  that the entanglement entropy is limited during the evaporation of black holes.
According to the prescription of quantum extremal surfaces the entanglement entropy is given, after renormalization\footnote{See \cite{Bombelli:1986rw,Srednicki:1993im} for renormalization of the entanglement entropy in four-dimensional case.}, by the formula 
 \be
\label{GEformula2}
\quad S(R) 
 = 
 \min \left\{\mathop{\mathrm{ext}}_{\cI}\left[
 \frac{\mathrm{Area}(\partial {\cI})}{4G} 
 + S_{\rm matter}(R\cup {\cI})
 \right]\right\} \,.
\ee
 Here $\cI$ is a region, called the "island", whose boundary area is denoted by $\mathrm{Area}(\partial \cI)$ and $S_{\rm matter}$
 is the von Neumann entropy  $S_{\rm vN}(R\cup I)$ of union of the island and the region $R$, $G$ is Newton's constant.
An extremization  on any possible island and then taking of
the configuration with the minimum entropy is assumed. 
As the  system evolves,  the form of island can change dynamically and  
different extreme surfaces can dominate at different times.  In several models this change of dominance provides the bound on $S(R)$ as time evolves \cite{Penington:2019npb,Almheiri:2019psf,Almheiri:2020cfm,Almheiri:2019yqk,Almheiri:2019qdq,Almheiri:2019psy,Chen:2019uhq,Penington:2019kki,
Marolf:2020xie,Hartman:2013qma,Faulkner:2013ana,Bousso:2019ykv,Balasubramanian:2020hfs}.
%,Hartman:2020swn,Faulkner:2013ana,Rozali:2019day,}.
\\

Although the prescription of the quantum extremal surface was  proposed in the
framework of holography \cite{Ryu:2006bv,Hubeny:2007xt}, the island rule is applicable to black holes in
more general theories.
The formula was confirmed for some two dimensional models \cite{Almheiri:2019hni,Almheiri:2019yqk}. 
For two dimensional gravity  the island rule has been derived by making use of replica trick \cite{Callan:1994py,Holzhey:1994we,Calabrese:2009qy}
 and the island contribution has been associated 
with replica wormholes \cite{Penington:2019kki,Almheiri:2019qdq}. The Page curve 
for  evaporating black holes in JT gravity has also been studied in 
\cite{Hollowood:2020}.  The wormhole configurations in JT gravity  have been discussed in \cite{Ageev:2019xii}.
For a further development in low dimensional case see
 \cite{Chen:2020jvn,
 Geng:2020fxl,
 Chen:2020uac,
Chen:2020hmv,
 Hernandez:2020nem,
 Colin-Ellerin:2020mva,
 Colin-Ellerin:2021jev,
 Balasubramanian:2021xcm}, 
 and references therein,  in particular, for  asymptotically flat two-dimensional  spacetime  see \cite{Gautason:2020tmk,Anegawa:2020ezn,Hartman:2020swn}.
The black holes in   four and higher dimensional  asymptotically flat spacetime have been considered in 
\cite{Almheiri:2019psy,Hashimoto:2020cas,Krishnan:2020,Alishahiha:2020qza,Matsuo:2020ypv}.
\\
 
Using the island formula \eqref{GEformula2}  for asymptotically flat eternal Schwarzschild black holes in four  spacetime dimension the saturation  at late time of  the entanglement entropy to the value
\be
 S_{\cI} 
 = 
 \frac{2\pi r_h^2}{G} + \frac{c}{6} \frac{b-r_h}{r_h}
 + \frac{c}{6} 
\log \frac{16r_h^3(b-r_h)^2}{G^2b} 
 \label{island2}
\ee
has been found in  \cite{Hashimoto:2020cas}.
Here $r_h$ is the gravitational radius of the Schwarzschild black hole, $r_h=2GM$, $M$ is the black hole mass, $c$ is the number of massless matter fields, $b$ defines  the boundaries of the entanglement regions in the right and the left wedges of the Schwarzschild geometry.
 The following inequalities are assumed: 
%\be \label{inequ}
$\sqrt{G}\ll r_h  \ll b$. 
%\ee
The entropy corresponding to non-island configuration dominates at small  time  and  increases with time linearly
\be \label{noisland}
 S _{n\cI}\simeq
 \frac{c}{6} \frac{t}{r_h}.  
\ee
Equalizing  the entropy without island with  the
entropy with  island under condition of dominating the first term in \eqref{island2}, one  estimates the  Page time \cite{Hashimoto:2020cas}
\be \label{tPage}
t_{Page}\sim \frac{6\pi r_h^3}{c \,G}. 
\ee
The above considerations, as noted in \cite{Hashimoto:2020cas},  do not resort neither to holographic correspondences, i.e.
embedding into higher-dimensional AdS spacetime, nor to coupling with an auxiliary system to absorb the radiation.
\\

In this paper we note that while applying the above estimates to the evaporation of a black hole, the second term $c\,b/6r_h$ in \eqref{island2}
 plays an important role and  becomes dominant for small $r_h$. Thus, the entropy starts to increase with decreasing mass of the black hole.
Or in other words,  although the inclusion of the island and the extremization of it's location leads to the time-independent entropy of the eternal black hole at late times, when the black hole evaporates to the Schwarzschild radius $ r_h \ll(c\,G\,b / 24 \pi ) ^ {1/3} $  entropy grows explosively over time.\\

 Let us discuss the origin for the appearance of the term $c\,b/6r_h$ in \eqref{island2} which is singular for small  $r_h=2GM$. We will see that the origin of this term lies in using of the Kruskal coordinates which are singular for small $G$.  The island formula  \eqref{GEformula2} has been derived in \cite{
Almheiri:2019yqk,
Almheiri:2020cfm} by using the replica method and expanding the quantum gravity path integral over gravitational constant $G$.  The semi-classical expansion  of the quantum gravity path integral over $G$ is different from the standard semi-classical expansion over the Planck constant because in the case of the $G$-expansion the background metric itself depends on $G$.
Moreover, the Schwarzschild solution in the Kruskal coordinates is singular for small $G$ (or small $M$), see a discussion of this question in \cite{Arefeva:2021byb}. As a result,                        the term $S_{\rm matter}(R\cup {\cI})$ is of the order $1/G$ for small $G$. Before  extremization the first term is of the same order and it  is given  by $2 \pi a^2/G$, 
here  $a$ is a  location of the island.
However the extremization procedure makes $a\approx r_h$ and $\,2\pi a^2/G=8\pi GM^2$, i.e.
this term becomes  subdominant  for small $G$ yielding leadership to the term $c\,b/(12G M)$
\footnote{In more detail, the entanglement entropy before extrimization as a function of $a$ has the form 
 $$S(a)
  =  
 \frac{2\pi a^2}{G} - \frac{c}{6} 
 \left(\frac{a}{r_h}  +\log \frac{a}{r_h}\right)+const. 
$$
One can see that $S(a)$ is  an increasing function for $a\geq r_h$ if $r_h^2>cG/12 \pi$. Therefore, the function has its minimum at $a=r_h$.}.
\\

\begin{figure}[t]
\centering
\begin{picture}(0,0)
\put(35,-5){$\Large{t_{Page}}$}
\put(75,-5){$\Large{t_{expl}}$}
\put(105,-5){$\Large{t_{blow}}$}
    \put(175,-5){$\Large{t_{Anti-Page}}$}
\put(225,-5){$\Large{t_{blow}}$}
\put(355,-5){$\Large{t_{blow}}$}
\end{picture}
\includegraphics[width=45mm]{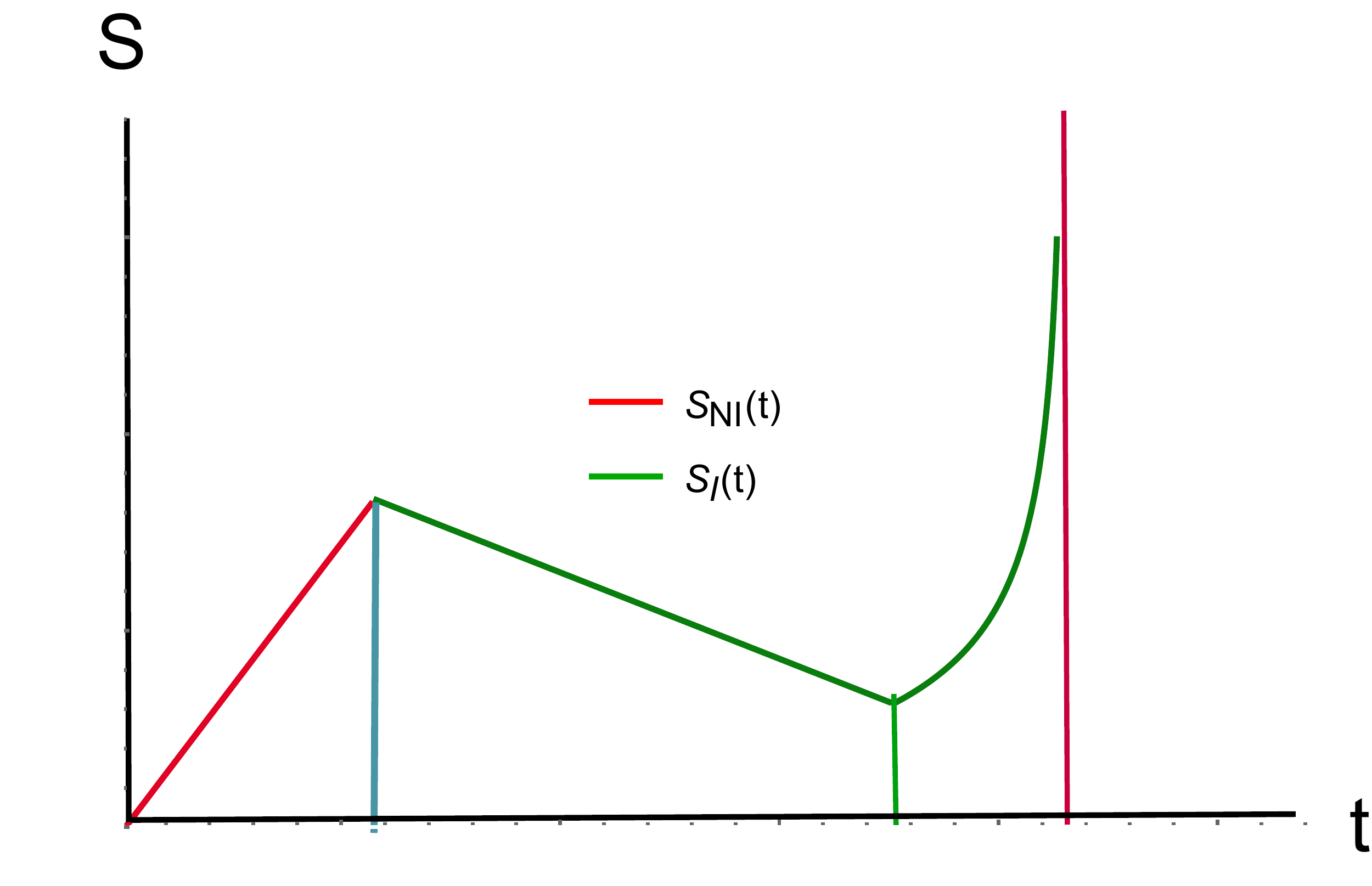}
\includegraphics[width=45mm]{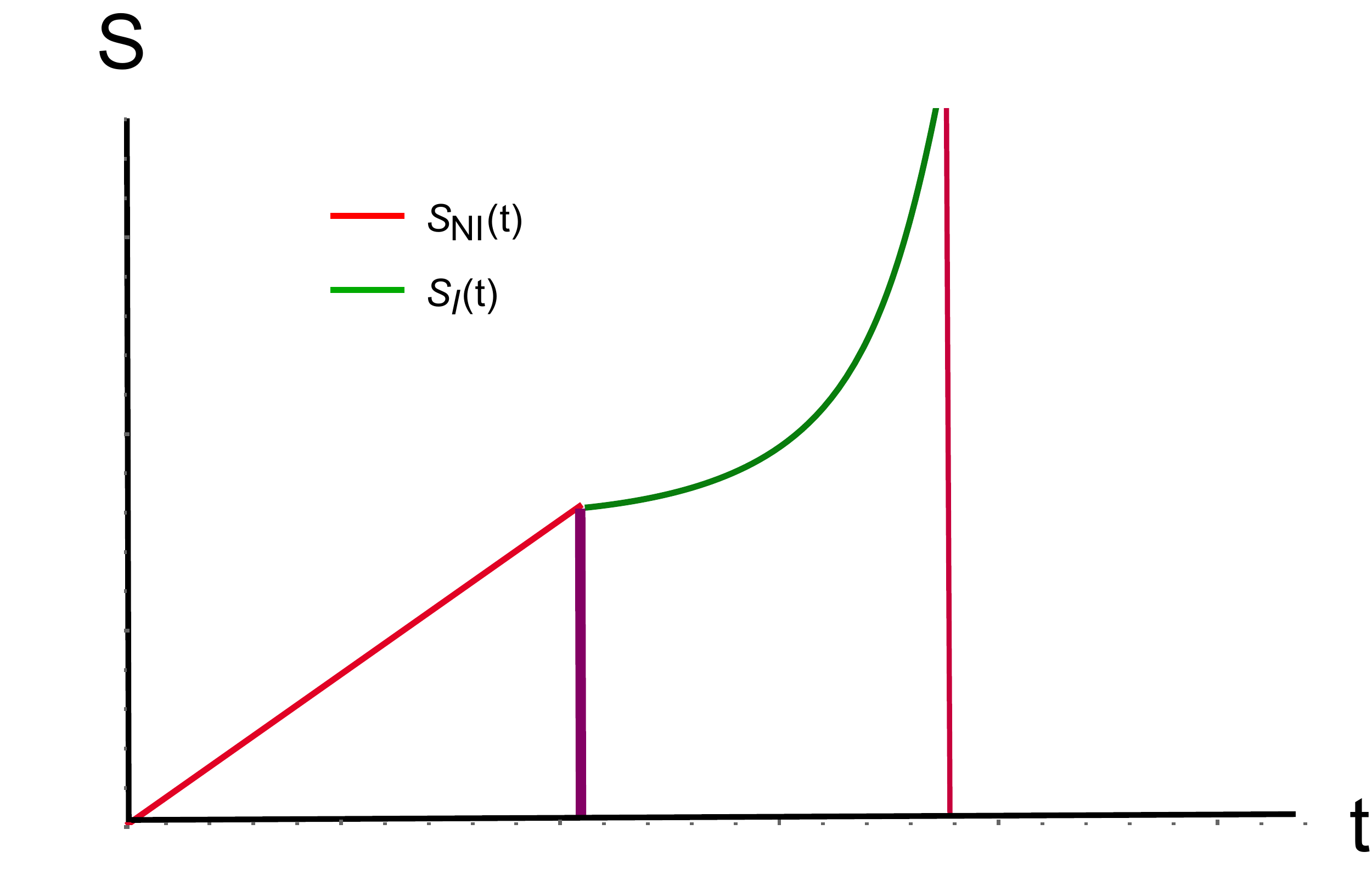}
\includegraphics[width=45mm]{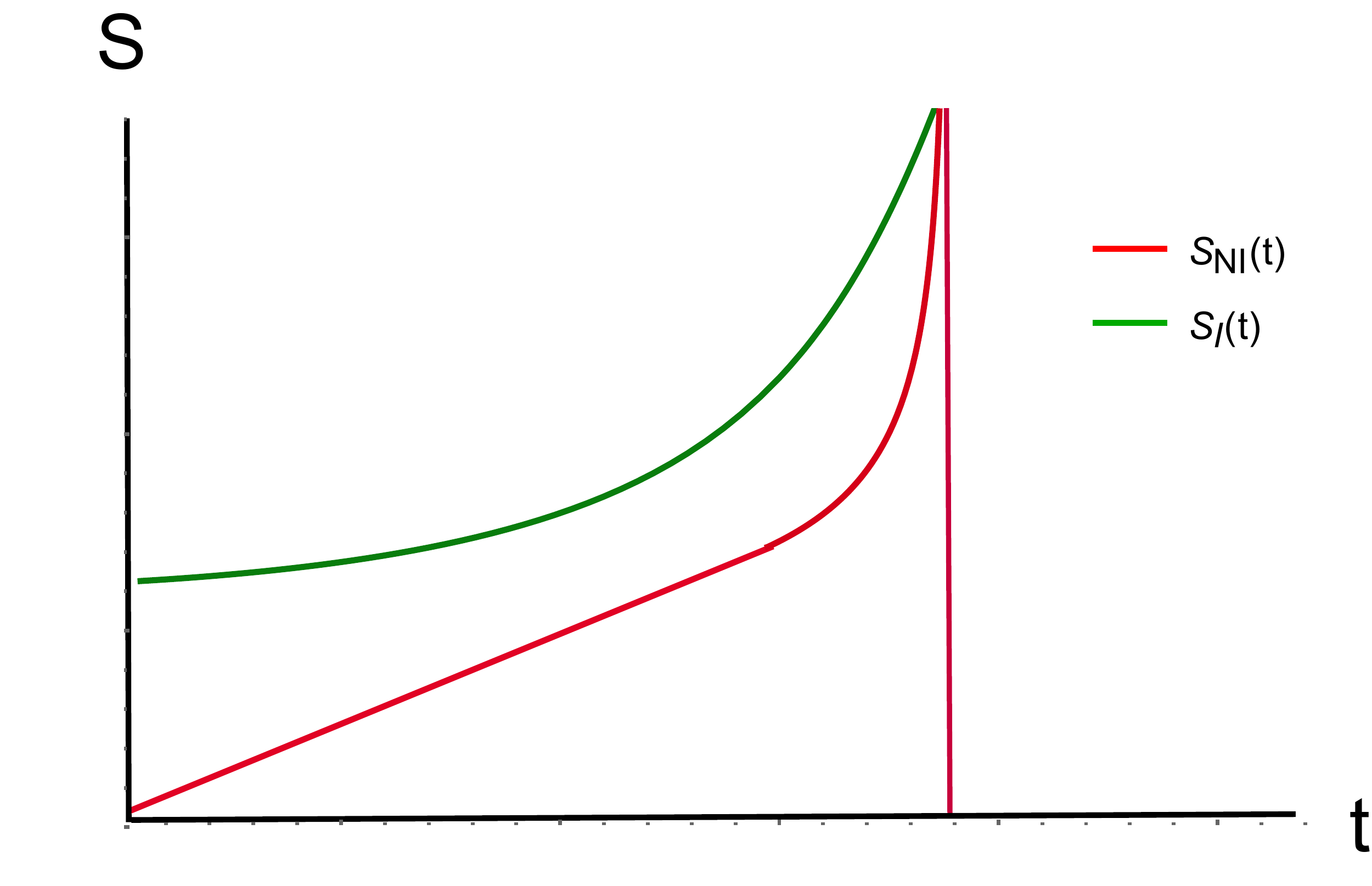}\\$\,$\\A)\hskip115pt B)\hskip115pt C)	\\\caption{The 
	 entropy with the island increases in the end of evaporation and becomes infinite at $t_{blow}$.  A) Increasing starts after some time of decreasing. 
	B)  There is no decreasing period.  The entropy increases after the intersection  point $t_{Anti-Page}$.  C) All the time $S_{n\cI}<S_{\cI}$. 
	%Label: fig:AntiPage.
	%{\bf Math.file: Page-curve-short.nb}
	}
	\label{fig:AntiPage}
\end{figure}
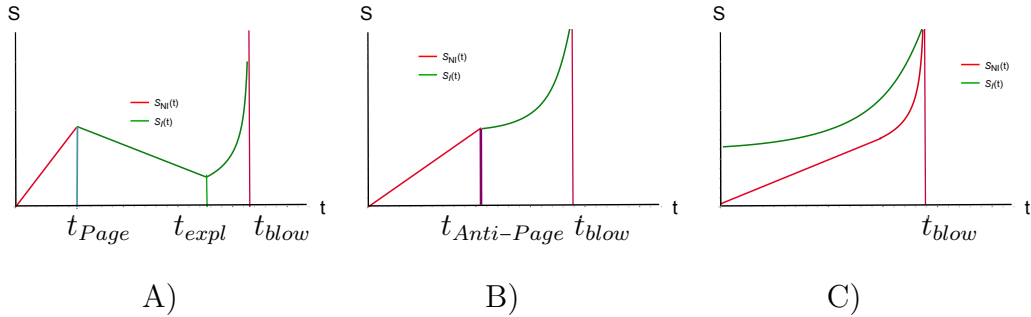

This is just an opposite behavior as compared with the estimation based on taking  only the first term $ 2\pi r_h^2/G$ into account in \eqref{island2} and  leading to decreasing of entropy to zero with decreasing of the black hole mass after the  Page time $t_ {Page}$ given  by \eqref{tPage}.
By taking  these two terms into account different scenarios  depending on the initial parameters of the evaporating black hole can be realized. 
\begin{itemize}
\item In  first scenario, 
before the Page time $ t_ {Page} $ the entropy increases. Then  the entropy decreases for some time, but then at the moment $ t_ {expl} $ an explosive growth of the entropy begins. One can say that time evolution follows the anti-Page curve.  This anti-Page  part of evolution ends with a blow up at the point $ t_ {blow} $, where the mass of the black hole is completely lost, see schematic plot in Fig.\ref{fig:AntiPage}.A. 
\item For another scenario,  the period of decreasing entropy
 disappears, see Fig.\ref{fig:AntiPage}.B.  In this case the initial increase of the  entropy, inherent in the configuration without an island,  at  moment $ t_ {expl} $ is replaced by
explosive behavior up to $ t_ {blow} $, typical for an island configuration with a low black hole mass.  This blow up continues   until complete evaporation at the moment $  t_ {blow} $. In this case, the Page point, where the increase of the entropy changes to decrease disappears,  altogether, and the anti-Page point, 
where the rate of growth of entropy changes, appears.

\item For more special case, Fig.\ref{fig:AntiPage}.C. the no-island configuration contribution  dominates all the time till total evaporation. 
\end{itemize}
We  also consider the question when the effect of an explosive growth of the entropy is behind the Planck scale.
 A  regularization of equation \eqref{island2} at $r_h\to 0$ is  discussed.

$$\,$$

The paper is organized as follows.  In the setup Section \ref{sec:Setup} we remind the formula for the entanglement entropy for configurations without and with an island for two sided eternal 4-dimensional  Schwarzschild  black hole obtained in \cite{Hashimoto:2020cas}.   In Section \ref{sec:timedep} we study  the  exchange of dominance between different configurations with an island and without an island for the evaporating black hole in details, especially in the end of evaporation.  Special attention is paid to the localization of the occurrence of quantum effects in this process. In Section \ref{sec:timedep-reg} we study a regularization of previous calculations that permits to consider the total evaporation of the black hole.

\section{Setup}\label{sec:Setup}
\subsection{Two-sided black hole}

\begin{figure}[h!]
\centering
\includegraphics[width=65mm]{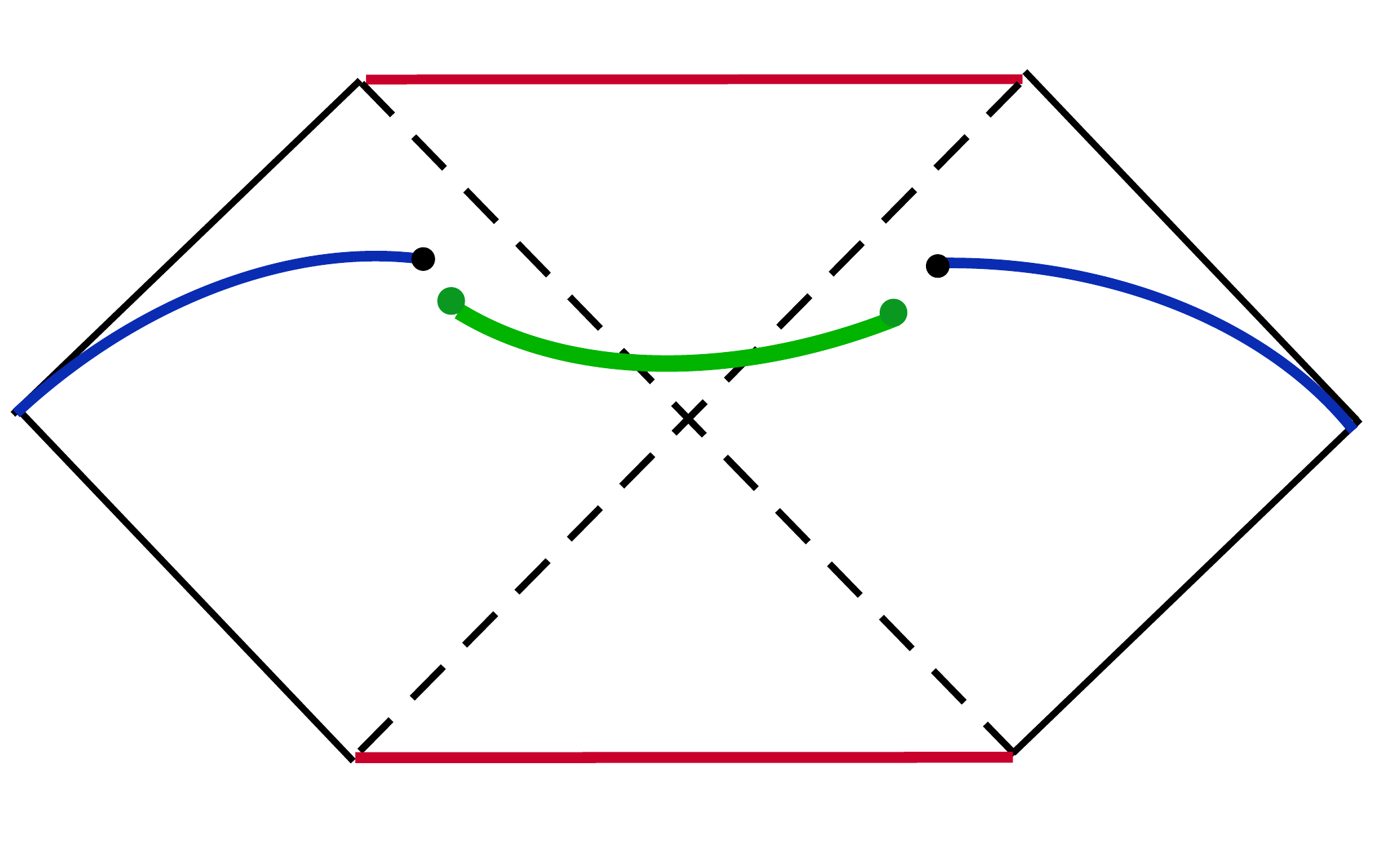}
\begin{picture}(0,0)
\put(-170,60){$\Large{R_{-}}$}
\put(-50,60){$\Large{R_{+}}$}
\put(-100,70){$\Large{\cI}$}
\put(-65,65){$\Large{b_{+}}$}
\put(-80,60){$\Large{a_{+}}$}
\put(-130,60){$\Large{a_{-}}$}
\put(-145,70){$\Large{b_{-}}$}
\end{picture}
\caption{The island configuration for two-sided black hole considered in \cite{Hashimoto:2020cas}. 
	% {\it NOT OUR PICTURE}
	}
	\label{fig:Island}
\end{figure}
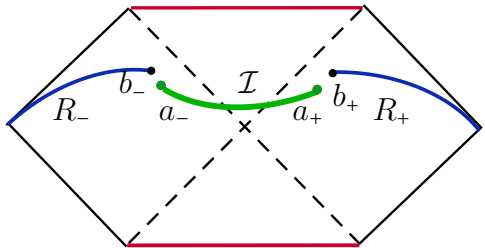
One of the  simple examples explicitly demonstrated  how an island can help to make  the bounded  entanglement entropy of the Hawking radiation is the  two-sided black hole \cite{Hashimoto:2020cas}.
The island formula for the generalized entropy consists of two parts
 \be\label{Sgen}
S_{gen}=S_{gr}+ S_{ vN}.
\ee 
the  gravity part $S_{gr}$ that is 
associated with a nontrivial quantum extremal surface,  island, and 
the matter (radiation) von Neumann entropy $S_{ vN}$.

One supposes that the radiation  is located at the union
of two regions $R_+$ and $R_-$,  which are located in the right and left wedges in the Penrose
diagram 
(see Fig.\ref{fig:Island})   near the null infinities, where the gravity 
is negligible.\\

The states that one considers are
maximally symmetric, so  the location of a possible island is fixed by its position in $(t,r)$ coordinates
and 
effectively one deals with  two-dimensional models
 with $(t,r)$ coordinates  (or deals only with s-modes) and 
computes the corresponding entanglement entropy between 
several entangling regions  using two dimensional answers 
\cite{Calabrese:2009qy}. \\

It is convenient to work  within  the Kruskal coordinates \cite{Wald} related with the Schwarzschild coordinates $t,r$ as 
\bea\label{UV}
U = - \sqrt{\frac{r-r_h}{r_h}}\,e^{-\frac{t-(r-r_h)}{2r_h}},\qquad V = \sqrt{\frac{r-r_h}{r_h}}\hspace{.5mm}e^{\frac{t+(r-r_h)}{2r_h}}
\eea
and by which the corresponding two-dimensional part of the Schwarzschild metric is
\bea\label{2D}
ds_{2-dim \,part \,Schw}^2 = - W^{-2} dU dV,\qquad W = \sqrt{\frac{r}{4r_h^3}} \,
e^{\frac{r-r_h}{2r_h}}\,.
\eea

%Entanglement entropy  of the Hawking radiation

For the configuration without islands the entanglement entropy $S_{n\cI}$ of the Haw\-king radiation 
is identified with that in the region $R = R_+\cup R_-$ and 
\begin{equation}
  S_{n\cI} = 
   \frac{c}{6} \log d(\ell_1,\ell_2) , 
  \label{S-nI}
\end{equation}
$d(\ell_1,\ell_2)$ is the geodesic distance between points $\ell_1$ and $\ell_2$ 
 given by 
\bea\label{dbpbm}
d(\ell_1,\ell_2)= \sqrt{\frac{\big(U(\ell_2)-U(\ell_1)\big)\big(V(\ell_1)-V(\ell_2)\big)}
{W(\ell_1)W(\ell_2)}}.
\eea  
Here points  $\ell_1$ and $\ell_2$ are located at $(t_b,b_+)$ and $(-t_b+i2 \pi r_h ,b_-)$.
The total entanglement entropy  for this configuration is given by  
\begin{equation}
 S _{n\cI}=
  %\frac{2\pi b^2}{G} + 
 \frac{c}{6} \log \left[\frac{16r_h^2(b-r_h)}{b}\cosh^2\frac{t_b}{2 r_h}\right].
\end{equation}
At $t_b \gg b \, (> r_h)$, the above result is 
approximated as
\begin{equation}
 S _{n\cI}\simeq
 \frac{c}{6} \frac{t_b}{r_h}\,  \qquad r_h=2 G M
 \label{noisland}
\end{equation}
and grows linearly in time. 
At the late times 
$ t\gg \frac{r_h^3}{cG}$ this entropy becomes much larger than the black hole entropy, and 
this contradicts with the finiteness of the von Neumann entropy for a finite-dimensional black hole system. 
In such a case an island is expected to emerge \cite{Hashimoto:2020cas}.  \\

For the configuration with the island, presented in Fig.\ref{fig:Island}, 
the entanglement entropy for the conformal matter is given by 
\bea
 S_{\text{matter}} &=& \frac{c}{3} \log \frac{d(a_+,a_-) d(b_+,b_-) d(a_+,b_+) d(a_-,b_-)}{d(a_+,b_-) d(a_-,b_+)}.
\eea
Locations of $a_\pm$ and $b_\pm$ are indicated in Fig.\ref{fig:Island} and $d(\ell_1,\ell_2)$ is given by \eqref{dbpbm}.Supposing  that $\sqrt{G}< r_h\ll b$ and the extremizing  of the total entropy about the location of the island,  one gets \cite{Hashimoto:2020cas} a unique island
location, that is $t_a=t_b$ and $a=r_h +r_h X^2(b,G,c)$, where $X^2<<1$.
At this configuration the total entanglement entropy is given by \eqref{island2}.
%\be
% S_{\cI} 
% = 
% \frac{2\pi r_h^2}{G} + \frac{c}{6} \frac{b-r_h}{r_h}
% + \frac{c}{6} 
%\log \frac{16r_h^3(b-r_h)^2}{G^2b} 
%\,. 
% \label{island}
%\ee
The main claim of \cite{Hashimoto:2020cas}, see also \cite{Alishahiha:2020qza}, is that  although at early times one has a linear growth, the island 
comes to rescue the unitarity at late times in agreement with the Page curve. Equalizing  the entropy without island $ S _{n\cI}$ with  the
entropy $ S _{\cI}$ with   island one estimates the  Page time given by \eqref{tPage}. The mutual information between the island and the region of collecting the Hawking radiation  has been considered in  \cite{Saha:2021ohr} and  the subsystem volume complexity  in this context has been considered in \cite{Bhattacharya:2021jrn}.\\

Note that the consideration in \cite{Hashimoto:2020cas}  and presented above is relied on the assumptions,  that  Hawking radiation has no
gravitational interaction and  the main contribution to the matter von Neumann entropy 
comes from the entanglement between s-wave modes of the quantum field, so one can reduce the theory into 
 two dimensional conformal field theory. 
 
 \subsection{Time dependence of the mass of the eva\-porating  black hole}\label{sec:TDEBH}

In four dimensions due to radiation the mass $M$ of the black hole is reduced as \cite{Page:1976df}
\be\label{Mt}
M(t)=\frac{r_0}{2 G} \left(1-\frac{24 \alpha  \,c \,G
   t}{r_0^3}\right)^{1/3},
   \ee
where $\alpha$ is a constant dependent on the spin of the radiating particle, $c$ is the number of massless matter fields
and  $r_0$ is the Schwarzschild radius at $t=0$. The semiclassical estimate of the black hole lifetime is 
\be\label{evapo}
t_{evaporate} = \frac{r_0^3}{24 c\,  \alpha G}.
\ee

\section{Time dependence of the entanglement entropy of the eva\-porating black hole}\label{sec:timedep}

In this section
using equation \eqref{island2} we analyze what happens, when the black hole is losing its mass. We consider this 
process  adiabatically  just supposing, that the dependence of the entanglement entropy on time is defined by dependence of mass of black hole on time $M(t)$ considered in \cite{Hashimoto:2020cas}. From formula \eqref{island2} one sees, that 
for small $r_h$, the term $cb/6r_h$ in \eqref{island2} dominates and in this case the entropy increases when
mass of the black hole goes to zero. We will show that just this increasing  leads  to so called  anti-Page time dependence of entropy of the system.\\

       \begin{figure}[h!]
\centering
\includegraphics[width=90mm]{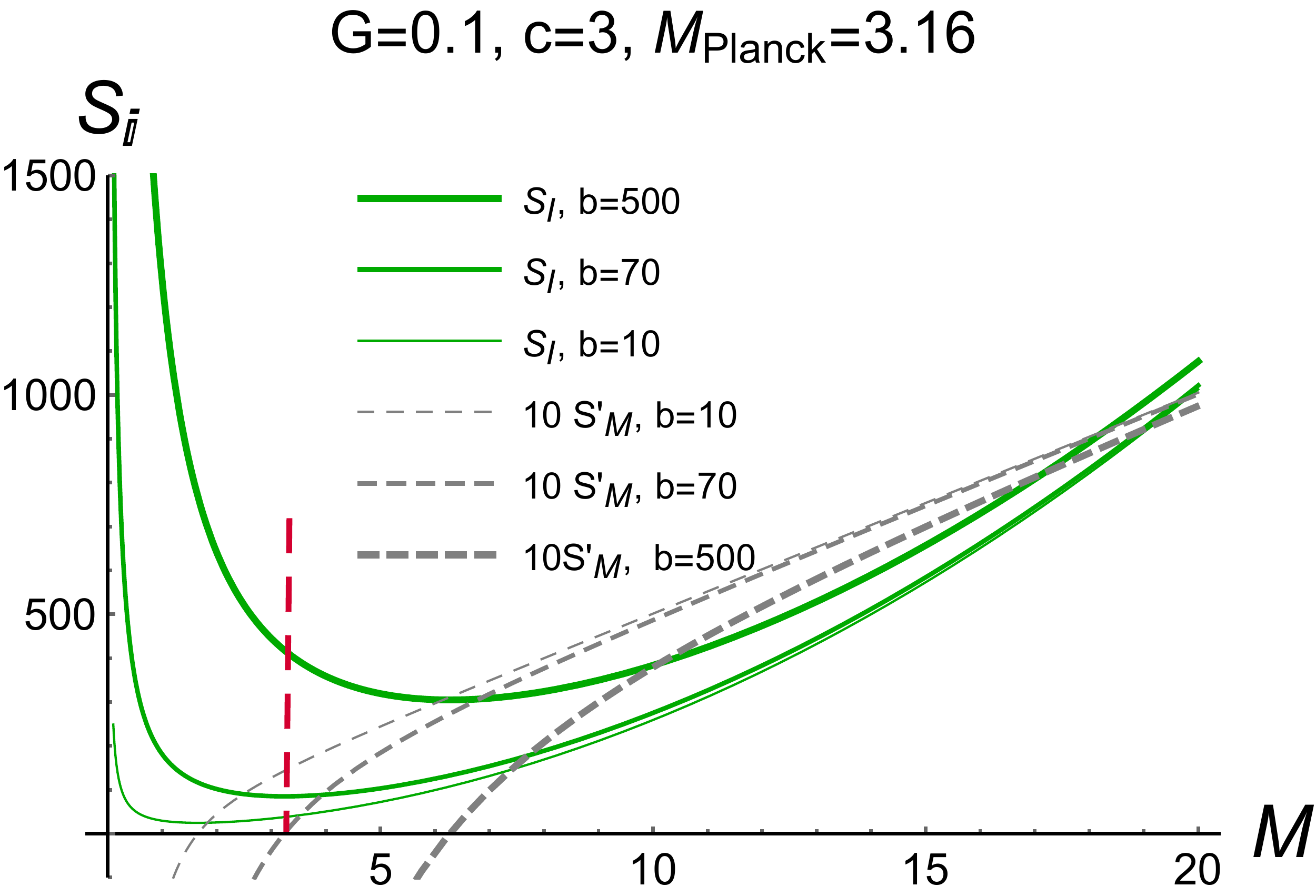}
\caption{Dependence of   $S_{Island}$ on $M$ (green lines). The dashed lines show $S'_{Island,M}$. The red dashed line shows the location of the Planck mass. 
%{\bf Math.file: Page-curve-short-Rusalev-04-10.nb}
}
	\label{fig:SM}
\end{figure}

The typical dependence of the entanglement entropy  \eqref{island2} on mass is presented in Fig.\ref{fig:SM}. We see that this dependence has a minimum located for large $b$, $b>r_h$ at 
\be\label{Mmin}
 M_{min}=\frac14\left(\frac{b c}{3
   \pi G^2 }\right)^{1/3}.
   \ee
This minimum can be realized outside of the Planck domain, i.e. 
\be M_{min}>M_{Planck} \simeq 1/\sqrt{G},\ee
 that corresponds 
in the used approximation to
\bea
  \sqrt{G}&<& \frac1{64}\frac{b c}{3 \pi }. \eea
  In all our plots  $c=3$, and 
therefore should be 
 $ \sqrt{G}< \frac{b }{64 \pi }.$%=0.004975 \,b$.
%  i.e.
%  \bea
%  G=0.001\quad  \mbox{we need }&\quad & b>6.358\nn\\
%   G=0.01 \quad  \mbox{we need }&\quad & b>20.10\nn\\
%    G=0.1 \quad  \mbox{we need }&\quad & b>63.5814\nn\\
%     G=1 \quad  \mbox{we need }&\quad & b>201.062\nn
%  \eea
%  ]]\\

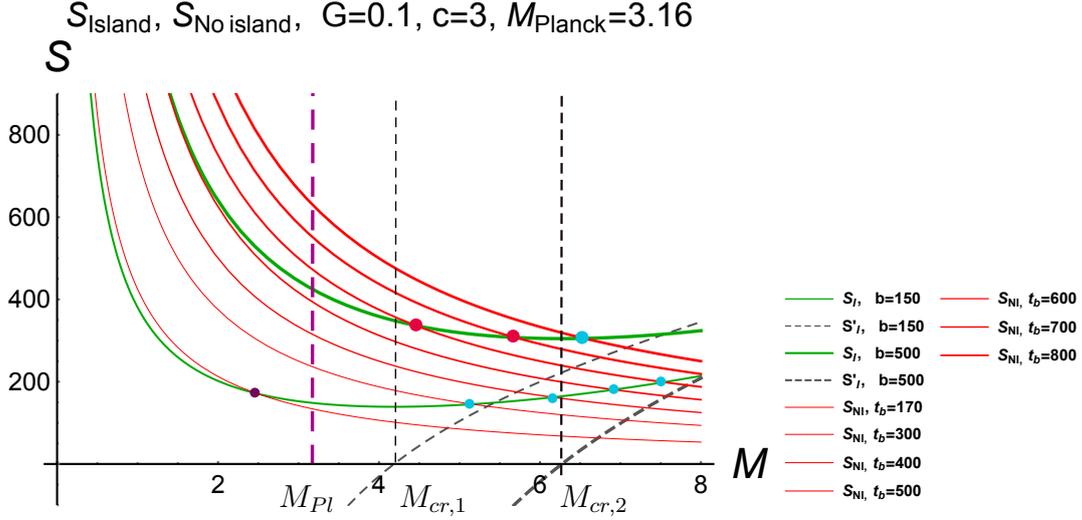
\begin{figure}[t!]
\centering
	\includegraphics[width=100mm]{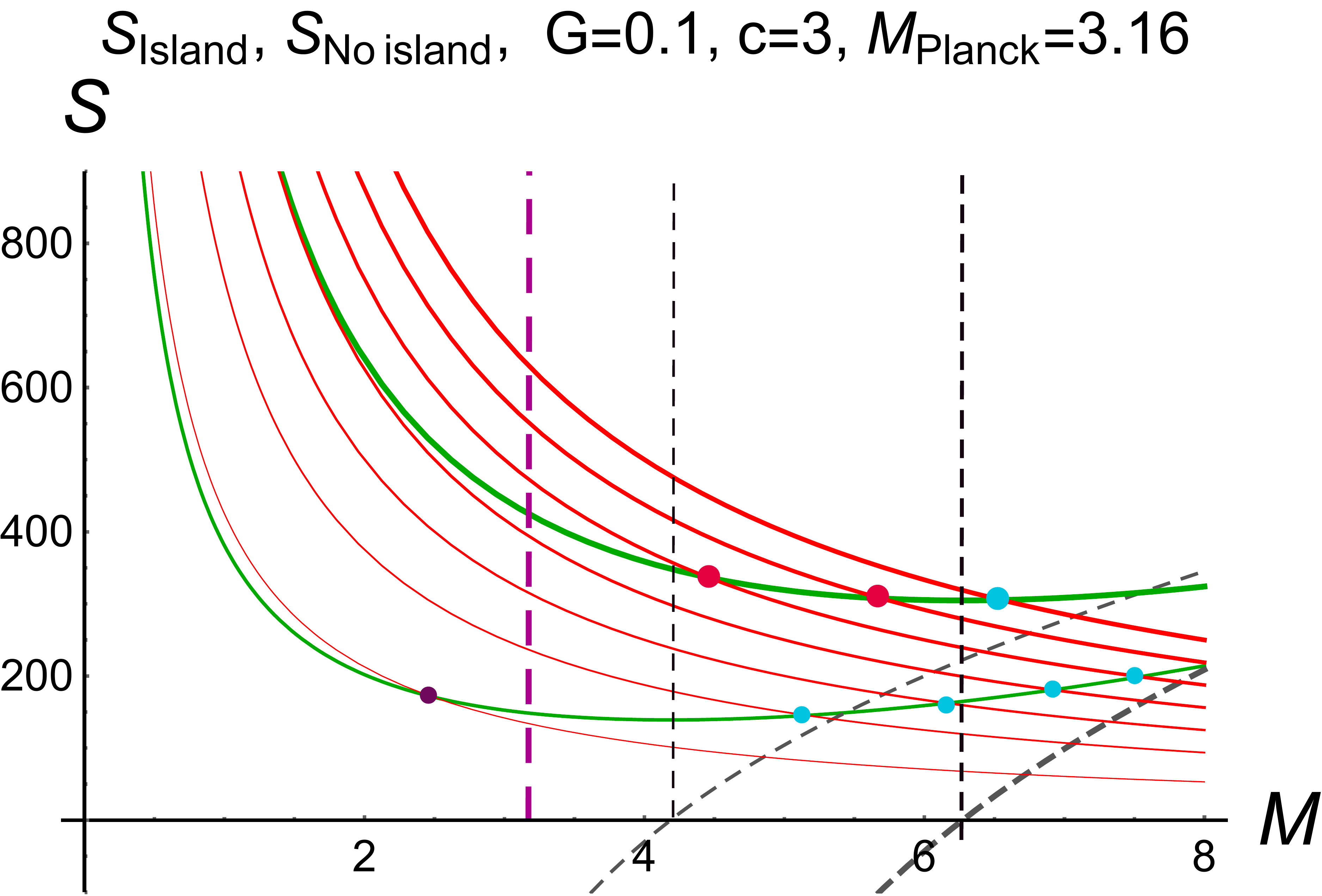}
	\includegraphics[width=40mm]{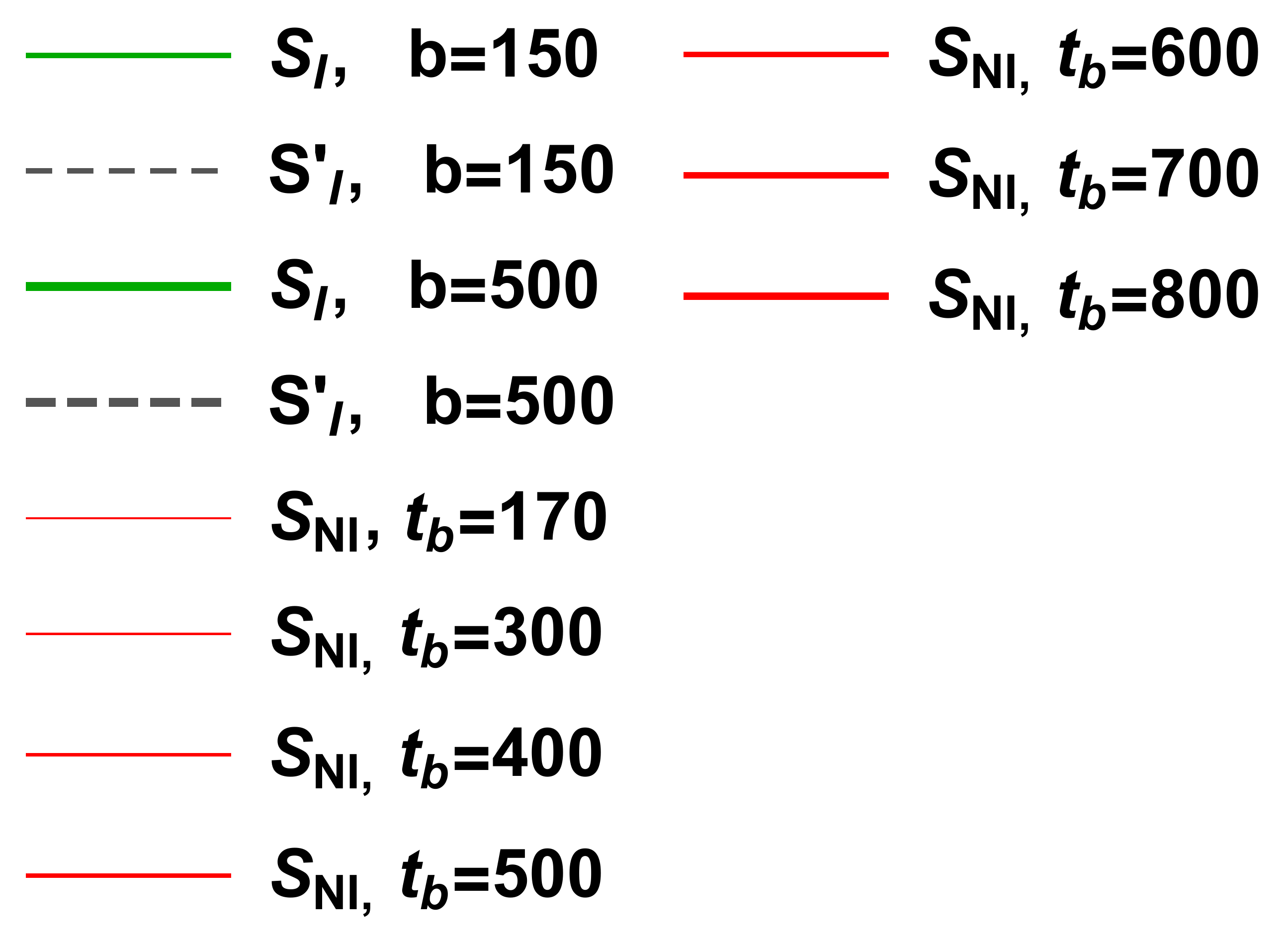}
		\begin{picture}(0,0)

\put(-200,0){$\Large{M_{cr,2}}$}
\put(-260,0){$\Large{M_{cr,1}}$}
\put(-305,0){$\Large{M_{Pl}}$}

\end{picture}
\caption{Dependence of the entropy  for the configuration with the island, $S_{\cI}$ (green line), and  the configuration  without the island, $S_{n\cI}$ (red lines), on $M$. Red lines of different thickness correspond to different $ t_b $
for the case  of $S_{n\cI}$.  Green lines of different thickness correspond to different $ b $ (here $b=150$ and $500$)
for the case  of $S_{\cI}$. The gray  lines show the derivative of  corresponding $S_{\cI}$ with respect to $M$. These derivatives are equal to zero at $M_{cr,1}$  and $M_{cr,2}$ for $b=150$ and $500$, respectively.
We see that $M_{cr,2}>M_{cr,2}$. The points show the intersections of the green lines with the red ones: darker cyan points show that at these points the increasing  of entropy with time changes to decreasing, while in the red points the increasing is preserved, only the slopes of the increase change. The magenta dot shows the intersection at masses less than the Planck mass.%{\bf Math.file: Page-curve-short-Rusalev-04-10.nb},
}
	\label{fig:SMm}
\end{figure}

One can compare the entropy with an island and the entropy for a configuration without an island, see Fig.\ref{fig:SMm}.
We see, that for a given mass of the black hole, after  some time, depending on the mass of the black hole, the entropy without  island (red lines with increasing thickness for increasing time) reaches the generalized entropy for the  configuration with the  island (the green line). This is true for both  increasing and decreasing branches of the entanglement  entropy with the island.
This observation gives the Page time, where two entropies are equalized. In the cases of cyan points indicated in Fig.\ref{fig:SMm}  the time dependence of the entropy for the eternal black  has a desirable form \cite{Hashimoto:2020cas}. In these cases 
dependence of  the Page time on mass of the black hole is given by
\be
t_{Page}= \frac{48\pi M^3}{c \,G^2}. \label{PagetimeM}
\ee
However, if the red line intersects the green line at $M$, that is less then corresponding $M_{cr}$ (two red points at
Fig.\ref{fig:SMm}), the increasing is preserved, only the slopes of the increase change. We call this point the  anti-Page
one, see the schematic plot presented in Fig.\ref{fig:AntiPage}.B.

\begin{figure}[h]
\centering
\begin{picture}(0,0)
\put(40,-10){$\Large{t_{Page}}$}
\put(180,-10){$\Large{t_{blow}}$}

\put(290,-5){$\Large{t_{expl}}$}
\put(370,-5){$\Large{t_{Pl}}$}
\put(405,-5){$\Large{t_{blow}}$}

\end{picture}
\includegraphics[width=70mm]{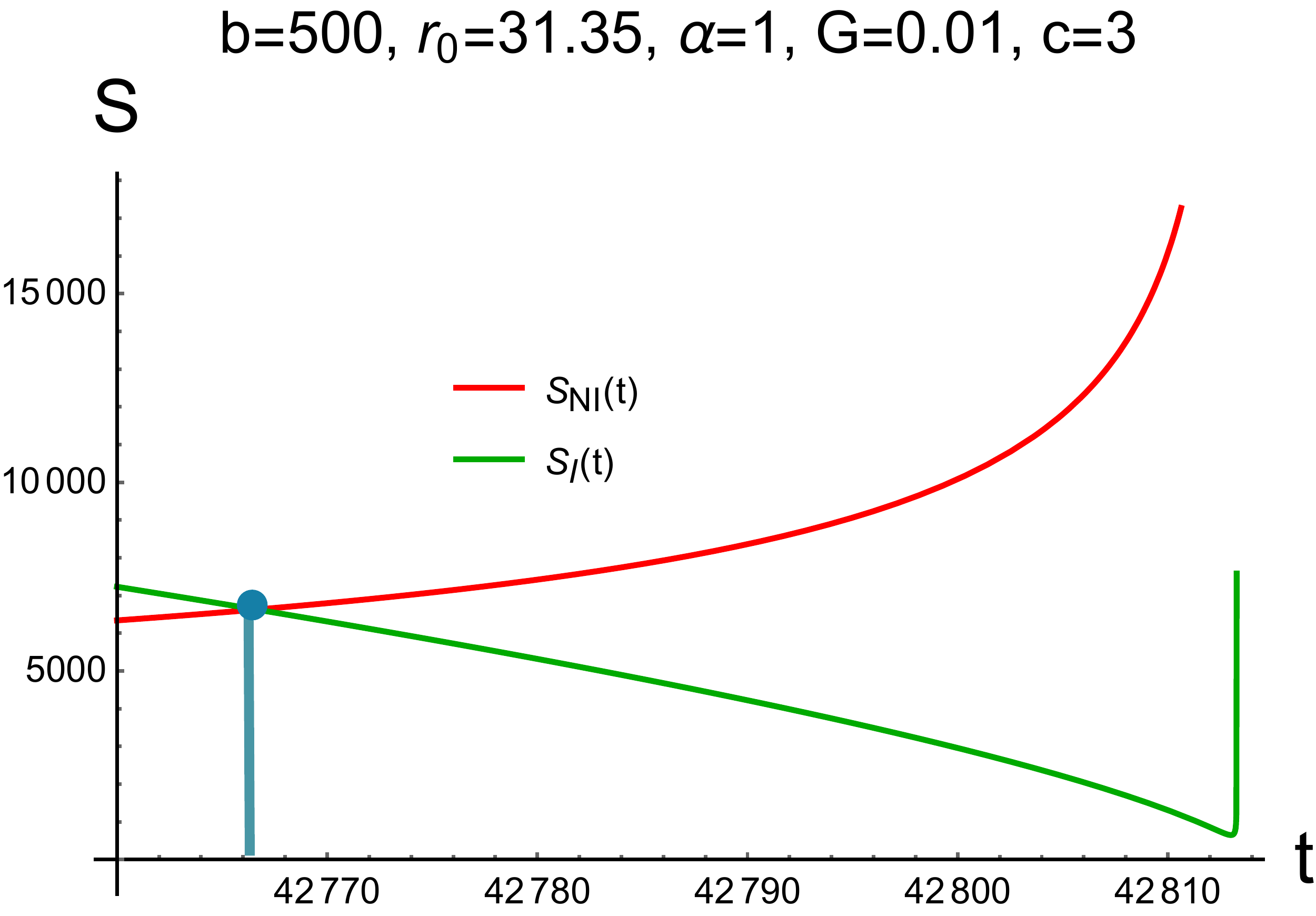}\qquad
	\includegraphics[width=70mm]{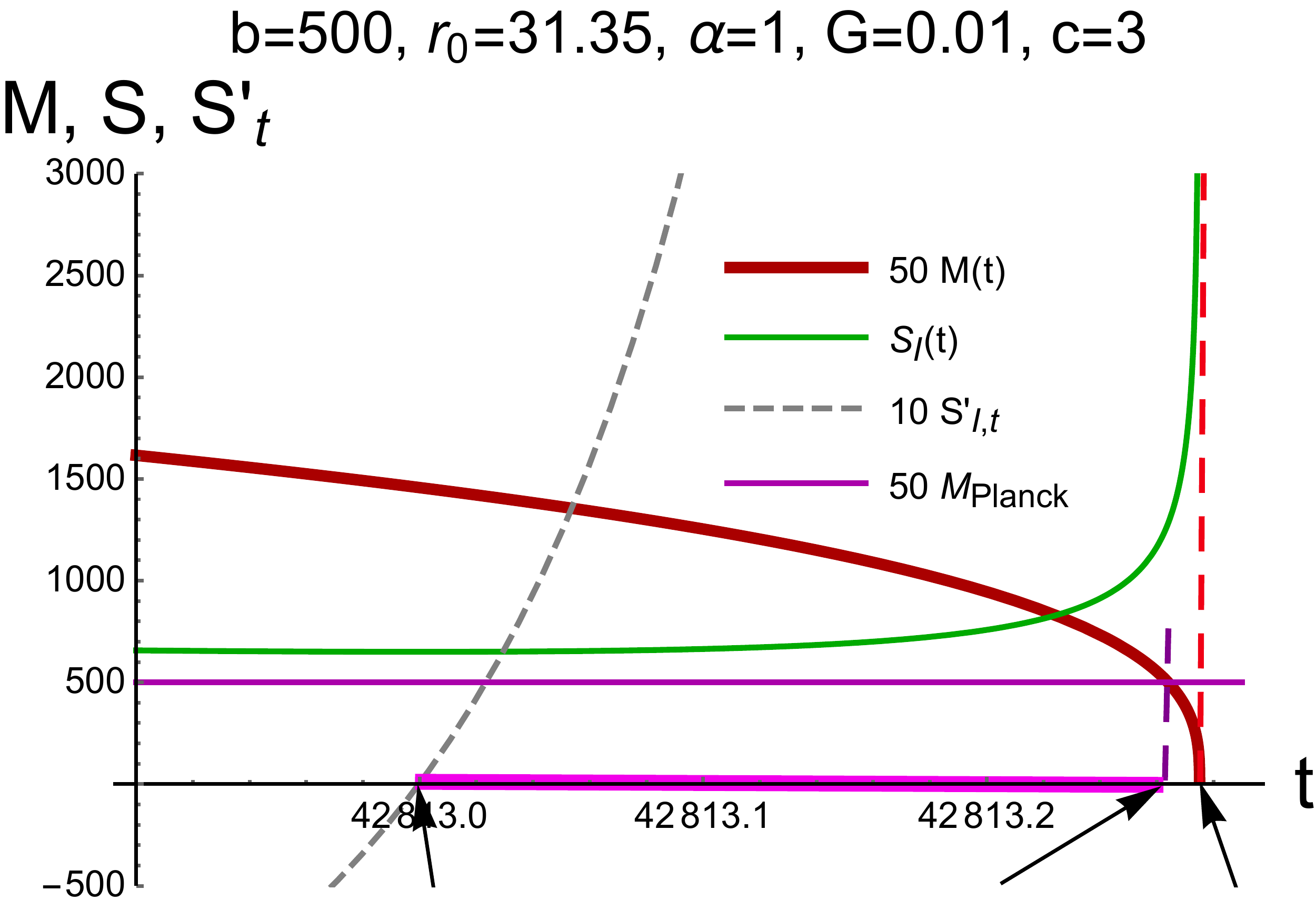}
	\\$\,$\\
	A\hskip150pt B\\
	\caption{ A) The green line shows the evolution of the configuration with the island, and the red one -- the configuration without  island. The cyan line indicates the Page time. 	B) Plot near the end of evaporation. The darker red line shows the decreasing  of mass of evaporating BH, the darker magenta line shows the Planck mass, corresponding to  $G=0.01$, i.e. $M_{Plank}=1/\sqrt{G}=10$, $t_{Pl}$ shows the time when $M(t)=M_{Pl}$, and 
	$t_{expl}$ and $t_{blow}$ show the times when the entropy starts to increase and increases to infinity, respectively.
	%{\bf Math.file: Page-curve-short-Rusalev-04-10.nb}
	}
	\label{fig:Mtb-cor}
\end{figure}

\begin{figure}[h!]
\centering
\begin{picture}(0,0)
\put(125,-10){$\Large{t_{Anti-Page}}$}
\put(215,-10){$\Large{t_{blow}}$}
\end{picture}
\includegraphics[width=90mm]{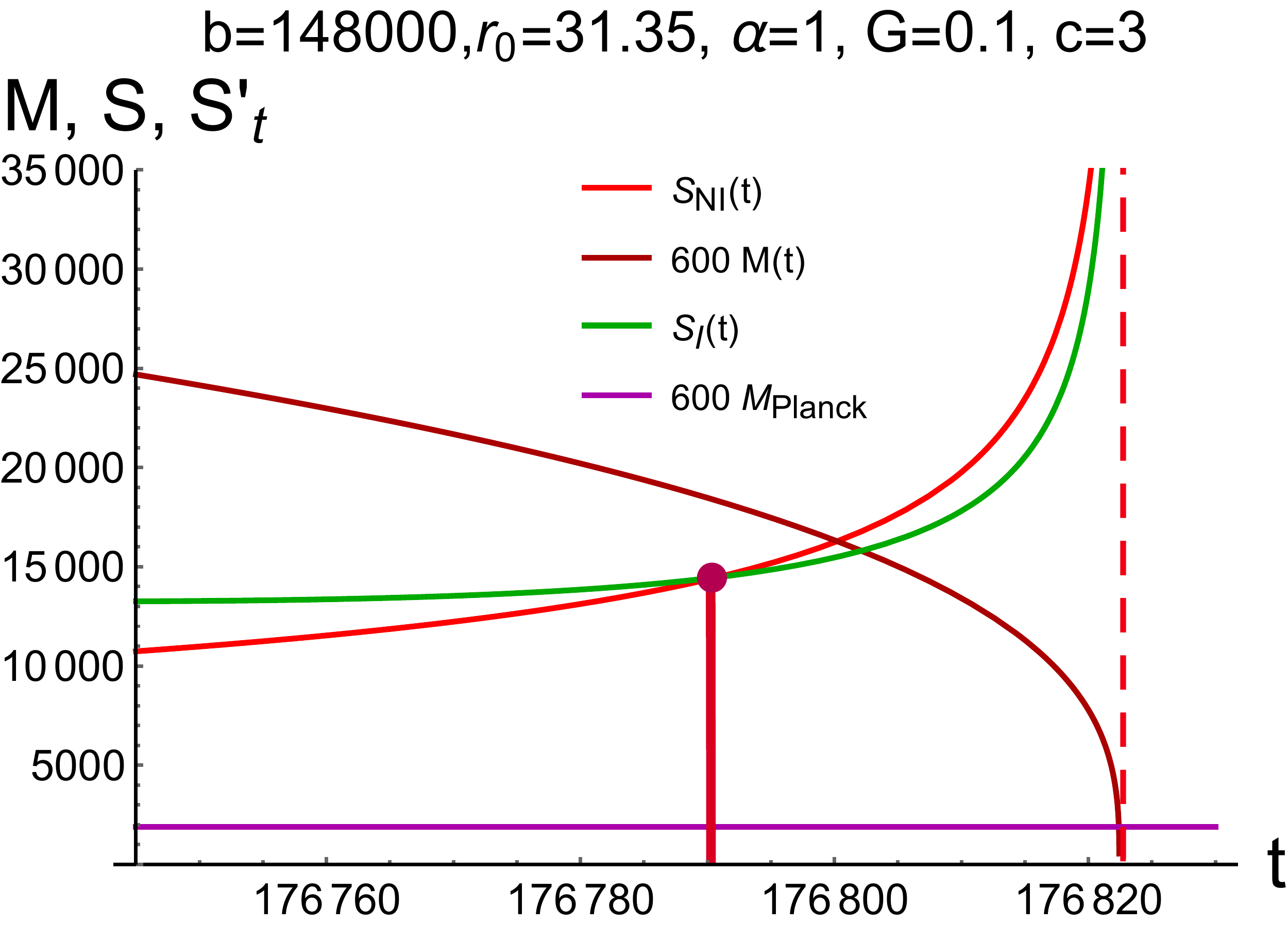}
\caption{The green line shows the evolution of the configuration with the island and the red one shows evolution of the  configuration without  island. The thick red line indicates the anti-Page time. 
	The darker red line shows the decreasing  of mass of evaporating BH, the darker magenta line shows the Planck mass, $M_{Planck}=1/\sqrt{G}$. Here   $G=0.1$ and $M_{Planck}=3.16$. 
	%{\bf Math.file: TR-EqualIncreaseEntropiesBeforePlankScale-28-10.nb}.
	}
	\label{fig:MtbG01}
\end{figure}

Now we would like to consider the modification of  the Page curve  for the black hole evaporation 
 following equation \eqref{Mt}. In plots Fig.\ref{fig:Mtb-cor}  and  Fig.\ref{fig:MtbG01}
the competition between two entropies is shown. The position of the Page time is indicated by the cyan line.
A new feature arises at the end of evaporation. As was mentioned in the Introduction, due to the presence of the term inverse of $r_h$ the island entropy starts to increase and blows up in the end of evaporation. This period of evolution may obscure behind the Planck scale or may not, depending on the parameters of the theory.

It may happen, that the growth of the entropy of the configuration with the island begins earlier than the two curves intersect,  see  
Fig.\ref{fig:MtbG01} and the schematic picture presented at  Fig.\ref{fig:AntiPage}. B.

\begin{figure}[h!]
\centering
\includegraphics[width=90mm]{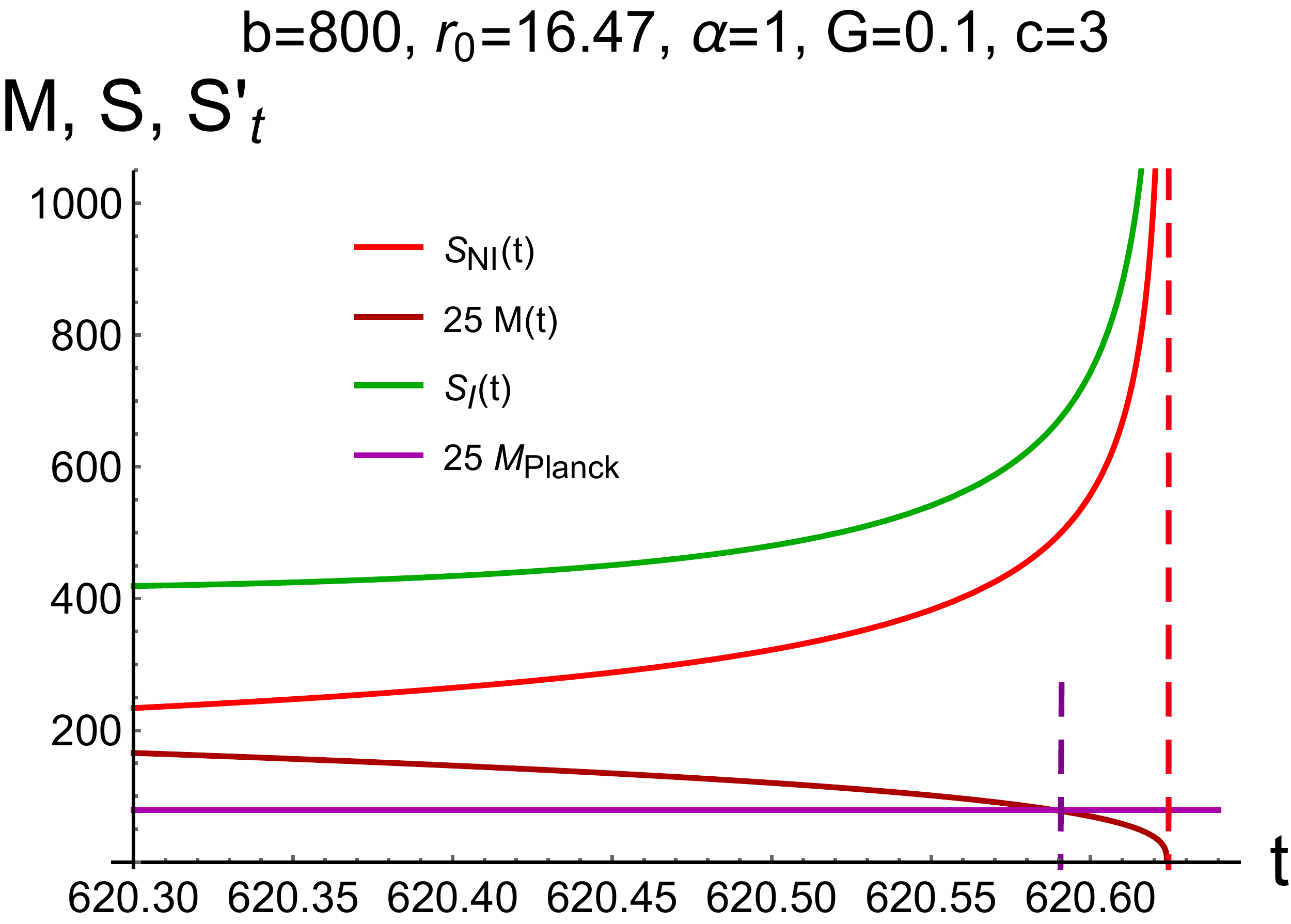}\\
\begin{picture}(0,0)
\put(75,0){$\Large{t_{Pl}}$}
\put(100,0){$\Large{t_{blow}}$}
\end{picture}
	\caption{ The green line shows the evolution of the configuration with the island and the red one configuration without  island. We see that there is no Page time at all, since the entropy for the configuration without   island lies below the entropy of the  configuration with the island in all periods of evaporation. The darker red line shows the decrease  of mass of the evaporating black hole, the magenta line shows the Planck mass, $M_{Planck}=1/\sqrt{G}$.
	%B) The zoomed plot shows the small region  near the end of evaporation. 
	The red line lies below the green one.  Here   $G=0.1$ and $M_{Planck}=3.16$. 
	%{\bf Math.file: Page-curve-short-Rusalev-04-10.nb}. %Label: fig:MtbG01-noPage
	}
	\label{fig:MtbG01-noPage}
\end{figure}

Moreover, the equalizing of two types of entropy may not appear before the complete  evaporation of the black hole, see  Fig.\ref{fig:MtbG01-noPage} and the schematic picture presented at  Fig.\ref{fig:AntiPage}. C.
We see at  Fig.\ref{fig:MtbG01-noPage} that entropy for the island configuration
begins to increase at some time and continues to exceed  the entropy of the non-island configuration for all the time until the end of evaporation.

Summarising this considerations, we get the plots schematically presented in Introduction, Fig.\ref{fig:AntiPage}.
Note that it is interesting to know if explosion regime is behind  the Planck scale. For this purpose  we indicate  the Planck  scale on all our plots. There are several cases:
\begin{itemize} 
\item  by the time $ t_ {expl} $ the black hole has not yet  evaporated to  the Planck mass,  see Fig.\ref{fig:Mtb-cor}. B;
\item by the time $ t_ {expl} $ the black hole has already evaporated to a mass less than the Planck mass, see  Fig.\ref{fig:MtbG01}.
\end{itemize} 

\begin{figure}[t]
\centering
\begin{picture}(0,0)
\put(55,-5){$\Large{t_{Page}}$}
\put(105,-5){$\Large{t_{Pl}}$}
\put(125,-5){$\Large{t_{expl}}$}
\put(150,-5){$\Large{t_{blow}}$}
\put(235,-5){$\Large{t_{Page}}$}
\put(290,-5){$\Large{t_{expl}}$}
\put(320,-5){$\Large{t_{Pl}}$}
\put(345,-5){$\Large{t_{blow}}$}
\put(60,-155){$\Large{t_{Pl}}$}
\put(75,-155){$\Large{t_{Anti-Page}}$}
\put(125,-155){$\Large{t_{blow}}$}
\put(230,-155){$\Large{t_{Anti-Page}}$}
\put(285,-155){$\Large{t_{Pl}}$}
\put(315,-155){$\Large{t_{blow}}$}
\end{picture}
	\includegraphics[width=65mm]{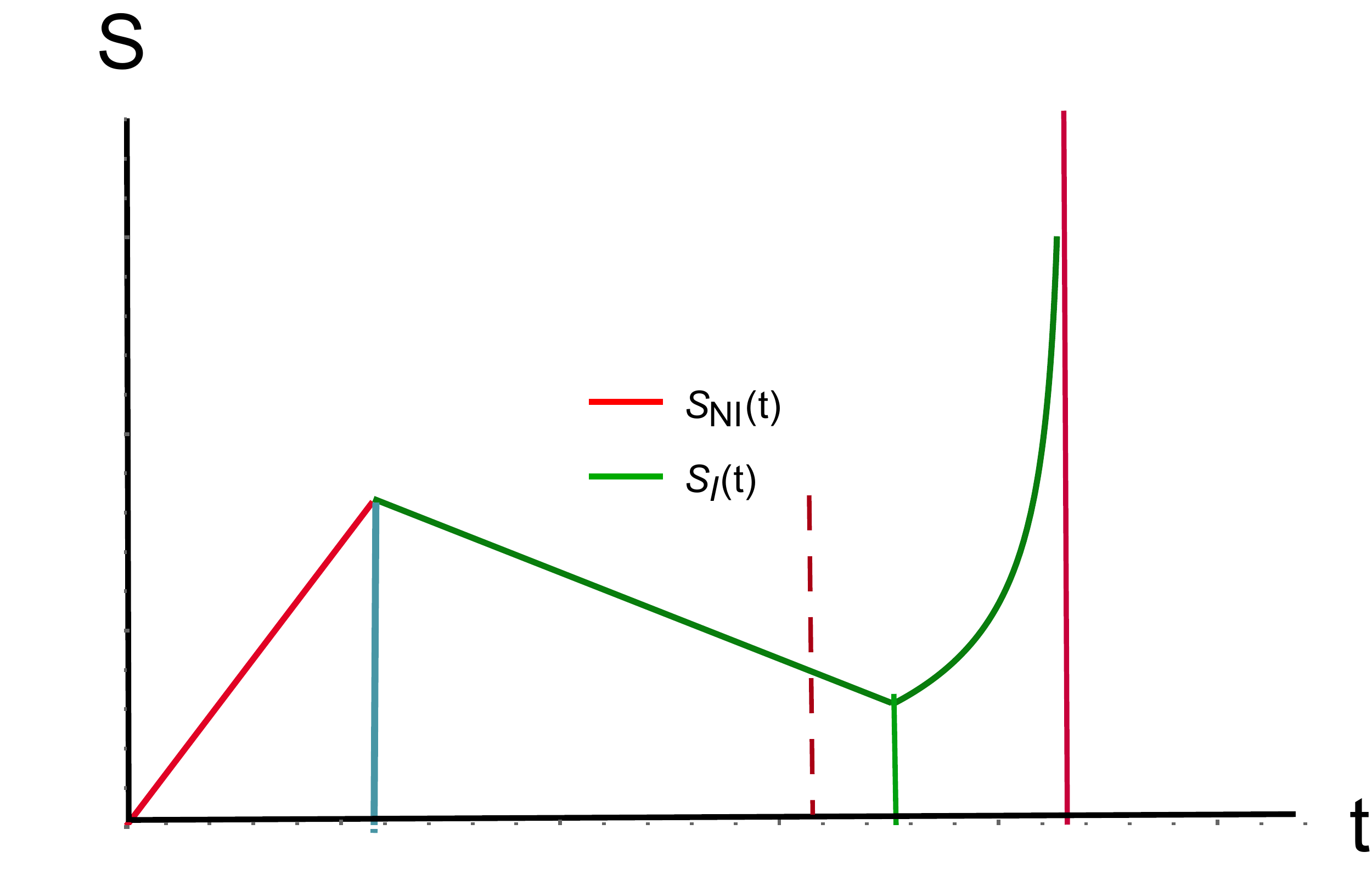}
	\includegraphics[width=65mm]{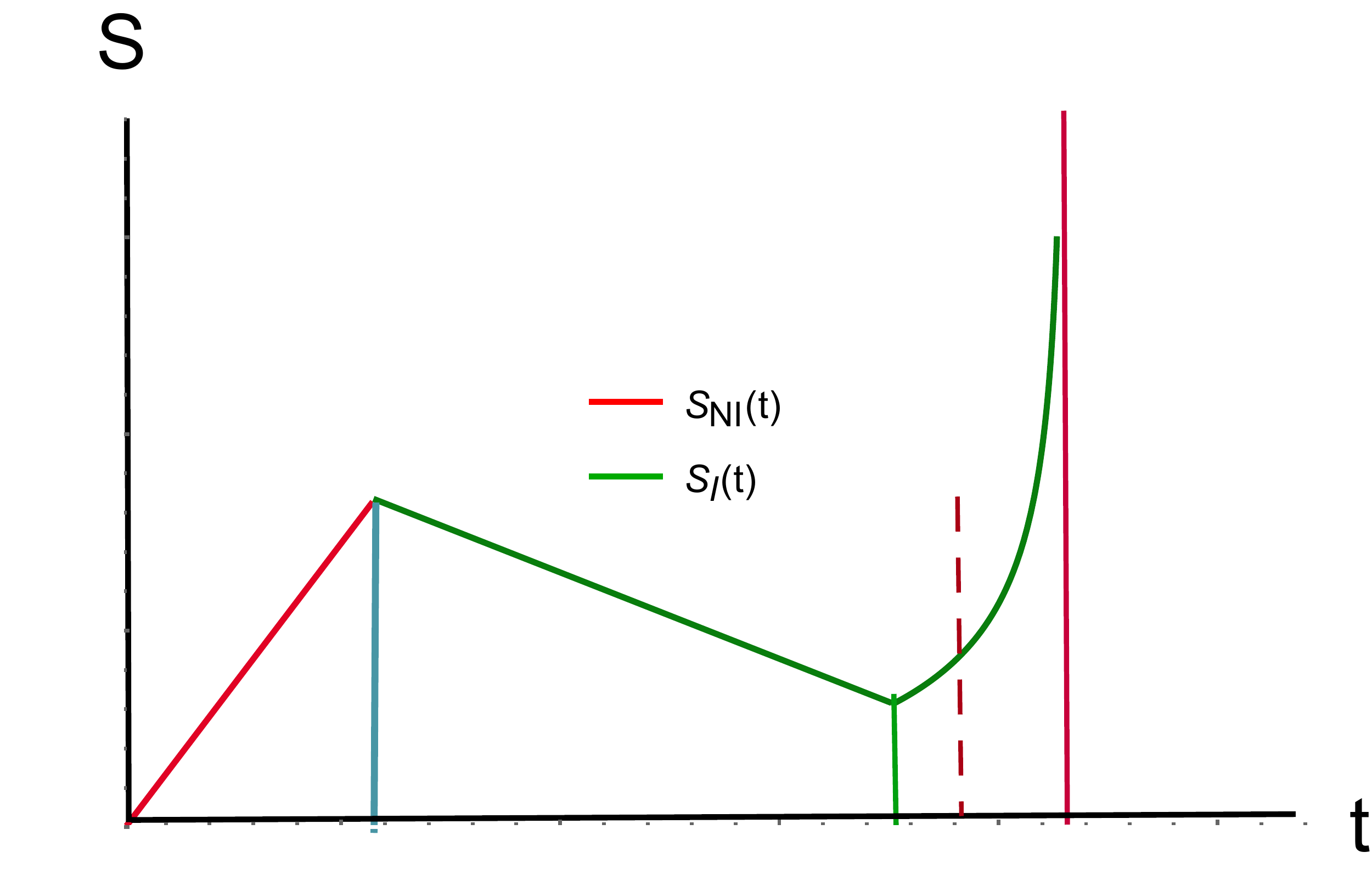}\\$\,$\\
	A\hskip150pt B\\
		\includegraphics[width=65mm]{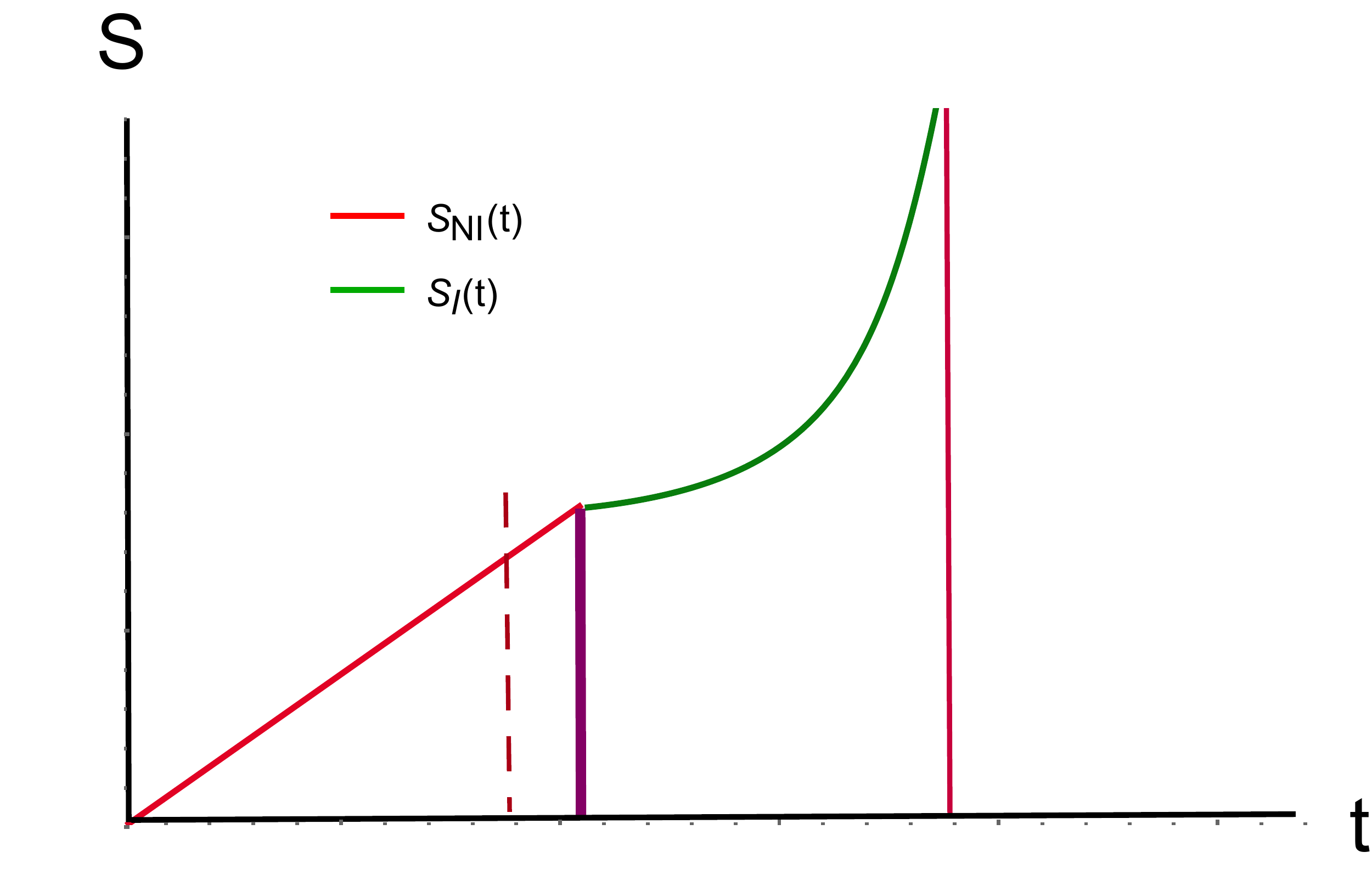}
		\includegraphics[width=65mm]{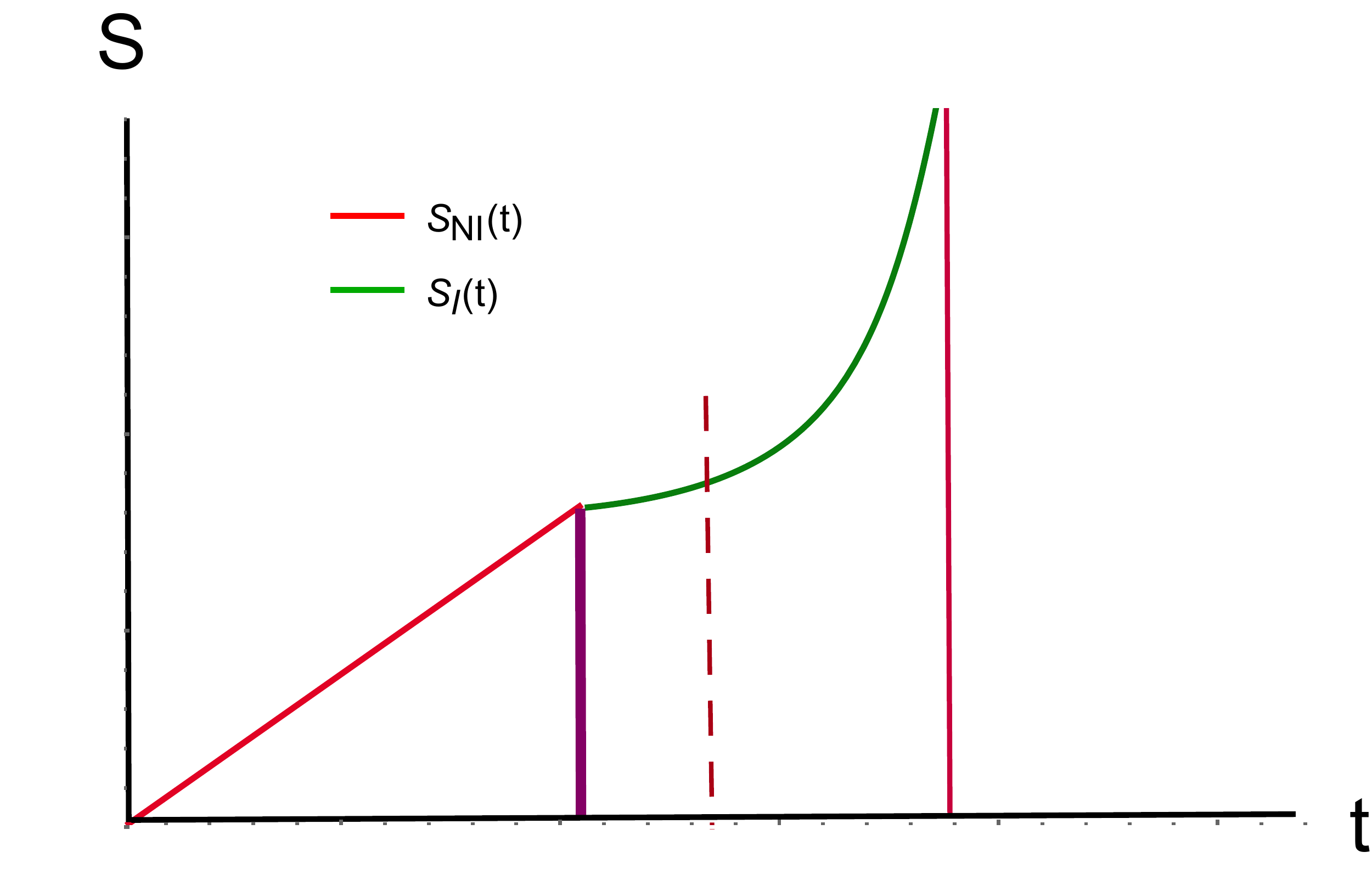}\\$\,$\\
	C\hskip150pt D\\
	\caption{
	The entropy  of configurations with the island increasing in the end of evaporation.  A) Increasing starts after some decreasing period and by the time $ t_ {expl} $ the black hole has evaporated to a mass less than the Planck mass.
	 B) Increasing starts after some decreasing period and at the moment $t_{expl}$ the black hole is still more that the   Planck mass.
	C)  There is no decreasing period and new increasing period starts at $t_{Anti-Page}$ and the mass of the BH in this time is less than  the Planck mass.
	D)  There is no decreasing period and new increasing period starts before the black hole becomes of  the Planck mass.  In all of these plots, $ t_ {Pl} $ indicates the time when the mass of the black hole becomes equal to the Planck mass.
	%Label: fig:AntiPage.
	%{\bf Math.file: Page-curve-short.nb}
	}
	\label{fig:AntiPage-Planck}
\end{figure}

In the first case we see the increasing region and therefore  the island does not solve the information problem.
In the second case one can say, that in  this increasing entropy region one has to modify the semi-classical approximation and include the quantum gravity corrections, since the point of the increasing $ t_ {expl} $ is out of the semiclassical approximation.
The same  concerns also to the case where the period  of decreasing is absent. In this case we have found the situation presented in Fig.\ref{fig:AntiPage-Planck}. C and D, i.e. the increasing point is obscured by quantum effects.

$$\,$$
\newpage

\section{Time dependence of the entanglement entropy of the eva\-porating black hole in thermal coordinates}
\label{sec:timedep-reg}

As we have seen in the previous section, time dependence of the entanglement entropy of the evaporating black hole in the end of evaporation has singular behaviour. This behaviour is related with singular behaviour of the Kruskal coordinates in the limit of the black hole mass $M\to 0$, see discussion of this problem in  \cite{Arefeva:2021byb}. One can remove this singularity using   thermal coordinates \cite{Arefeva:2021byb}, that provide a  regularization of the Kruskal coordinates near $M=0$. These coordinates are introduced as
\bea\label{UmK}
\Um&=&- e^{-\frac{t - ( r- r_h) }{B}} \left(\frac{r-r_h}{r_h}\right)^{\frac{r_h}{B}},
\\
\label{VmK}
\Vm&=&e^{\frac{t + ( r- r_h) }{B}} \left(\frac{r-r_h}{r_h}\right)^{\frac{r_h}{B}},
\eea
where
\be\label{B}
B=(4M+\mu)G, 
\ee
and $\mu$ is a positive constant. In these coordinates the standard  Schwarzschild solution has the temperature 
$T=1/2\pi B$.  Two dimensional part of the Schwarzschild 
metric is 
\bea\label{2Dth}
ds_{2-dim \,part \,Schw}^2 &=&
\frac{\,d\Um d\Vm}{\Wm^2},\qquad
\Wm=\frac1{B}\left(\frac{\Um \Vm}{ (1 - \frac{2 G M}{r}) }\right)^{1/2},\eea
compare with \eqref{2D}.
For the configuration without islands, by analogy with \eqref{S-nI}, the entanglement entropy of the Hawking radiation 
is identified with that in the region $R = R_+\cup R_-$ and it is
\begin{equation}
  S_{n\cI} = 
   \frac{c}{6} \log \cD (\ell _1,\ell_2) , 
  \label{S-nI}
\end{equation}
where $d(\ell_1,\ell_2)$ is the geodesic distance between points $\ell_1$ and $\ell_2$
 given by 
\bea\label{Dbpbm}
{\cal D} (\ell_1,\ell_2)& =& \sqrt{\frac{\left(\Um(\ell_2) - \Um(\ell_1)\right)\left(\Vm(\ell_1) - \Vm(\ell_2)\right)}{\Wm(\ell_1)\Wm(\ell_2)}}.
\eea  
Here points  $\ell_1$ and $\ell_2$ are located at $(t_b,b_+)$ and $(-t_b+i \pi B ,b_-)$.
The regularized  entanglement entropy  for the  configuration presented in Fig.\ref{fig:Island}  is given by  
\be
 S _{\cI,reg}= \frac{2\pi a^2}{G} 
 + \frac{c}{3} \log \cL_{reg}(a,b,t_a,t_,b,r_h,\mu), 
\ee
where 
\bea
\cL_{reg}(a,b,t_a,t_,b,r_h,\mu) ^2=\cR_{reg}(a,b,t_a,t_,b,r_h,\mu) =\frac{\cD(a_+,a_-) \cD(b_+,b_-) \cD(a_+,b_+) \cD(a_-,b_-)}{\cD(a_+,b_-) \cD(a_-,b_+)}\nn\label{cR}.\eea
\bea
\cL_{reg} =\frac{\cD(a_+,a_-) \cD(b_+,b_-) \cD(a_+,b_+) \cD(a_-,b_-)}{\cD(a_+,b_-) \cD(a_-,b_+)}\nn\label{cR}.\eea

The explicit expression for $\cR_{reg}$ is given by 
   \bea
   &&\cR_{reg}=\frac{16 B^4 (a-r_h)(b-r_h) \cosh
   ^2\left(\frac{t_a}{B}\right) \cosh
   ^2\left(\frac{t_b}{B}\right)
   \left(e^{\frac{a-b}{B}}
   \left(\frac{a-r_h}{b-r_h}\right)^{\frac{r_h}{B}}+e^{\frac{b-a}{B}}
   \left(\frac{b-r_h}{a-r_h}\right)^{\frac{r_h}{B}}-2 \cosh
   \left(\frac{t_a-t_b}{B}\right)\right)^2}{a b G^2
   \left(e^{\frac{a-b}{B}}
   \left(\frac{a-r_h}{b-r_h}\right)^{\frac{r_h}{B}}+e^{\frac{b-a}{B}}
   \left(\frac{b-r_h}{a-r_h}\right)^{\frac{r_h}{B}}+2 \cosh
   \left(\frac{t_a+t_b}{B}\right)\right)^2}\textcolor{red}{.}\nn\\
   \label{AnsRc}
   \eea
For $\mu=0$ this expression reproduces the corresponding expression from \cite{Hashimoto:2020cas}
and as we have seen in the previous Section \ref{sec:timedep} does not admit the finite limit for  $r_h=0$.

%To get the $r_h\to 0$  limit of the regularized entanglement entropy $S_{\cI,reg}$ we take the limit  $r_h\to 0$  in expression \eqref{AnsRc}  and then find the  regularized entanglement entropy for zero mass  by extremising $S_0=S_{\cI,reg}(a,b,t_a,t_,b,0,\mu)$
%on $a$  and $t_a$. 

To get the limit $r_h \to 0 $ of the regularized entanglement entropy $S_{\cI, reg}$, we take the limit $r_h \to 0 $ in the expression \eqref {AnsRc}, and then find the regularized entanglement entropy for zero mass by finding extremum of the expression $ S _ {reg,0} $  varying over $ a $ and $ t_a $. 
$S_{reg,0}$ is given as 
\be
S_{reg,0}=\lim_{r_h\to 0} S_{reg,rh}=\frac{2\pi a^2}{G}+\frac {c}{6}\log \cR_{reg,0},
\ee
where
   \bea
   \cR_{reg,0}&\equiv&\cR_{reg}\Big|_{r_h\to 0}=\frac{16 B_0^4  \cosh
   ^2\left(\frac{t_a}{B_0}\right) \cosh
   ^2\left(\frac{t_b}{B_0}\right)
   \left(e^{\frac{a-b}{B_0}}
   +e^{\frac{b-a}{B_0}}-2 \cosh
   \left(\frac{t_a-t_b}{B_0}\right)\right)^2}{ G^2
   \left(e^{\frac{a-b}{B_0}}
   +e^{\frac{b-a}{B_0}}
   +2 \cosh
   \left(\frac{t_a+t_b}{B_0}\right)\right)^2},\nn\\
   B_0&=&\mu \,G. 
   \eea
   Extremizing on $t_a$ at large $t_b,t_b\to \infty$ we get that $t_a=t_b$ and at large $t_b,t_b$
    \bea
   &&\cR_{reg,0}\approx\frac{4B_0^4 }{ G^2}    \left(\cosh
   \left(\frac{a-b}{B_0}\right)- 1\right)^2.\nn
   \eea
   Assuming $a,b>B_0$ we have   
   \bea
   S_{0,reg,apr}& \simeq  &\frac{2\pi a^2}{G}+\frac {c}{6}\log \left[4 G^2 \mu^4    \cosh
   \left(\frac{a-b}{B_0}-1\right)^2\right]\textcolor{red}{.}\label{Sapr} \eea
    \bea
   S_{\cI,reg}\Big|_{M=0}& \simeq  &\frac{2\pi a^2}{G}+\frac {c}{6}\log \left[4 G^2 \mu^4    \cosh
   \left(\frac{a-b}{\mu \,G}-1\right)^2\right]\textcolor{red}{.}\label{Sapr} \eea

\begin{figure}[h!]

\includegraphics[width=65mm]{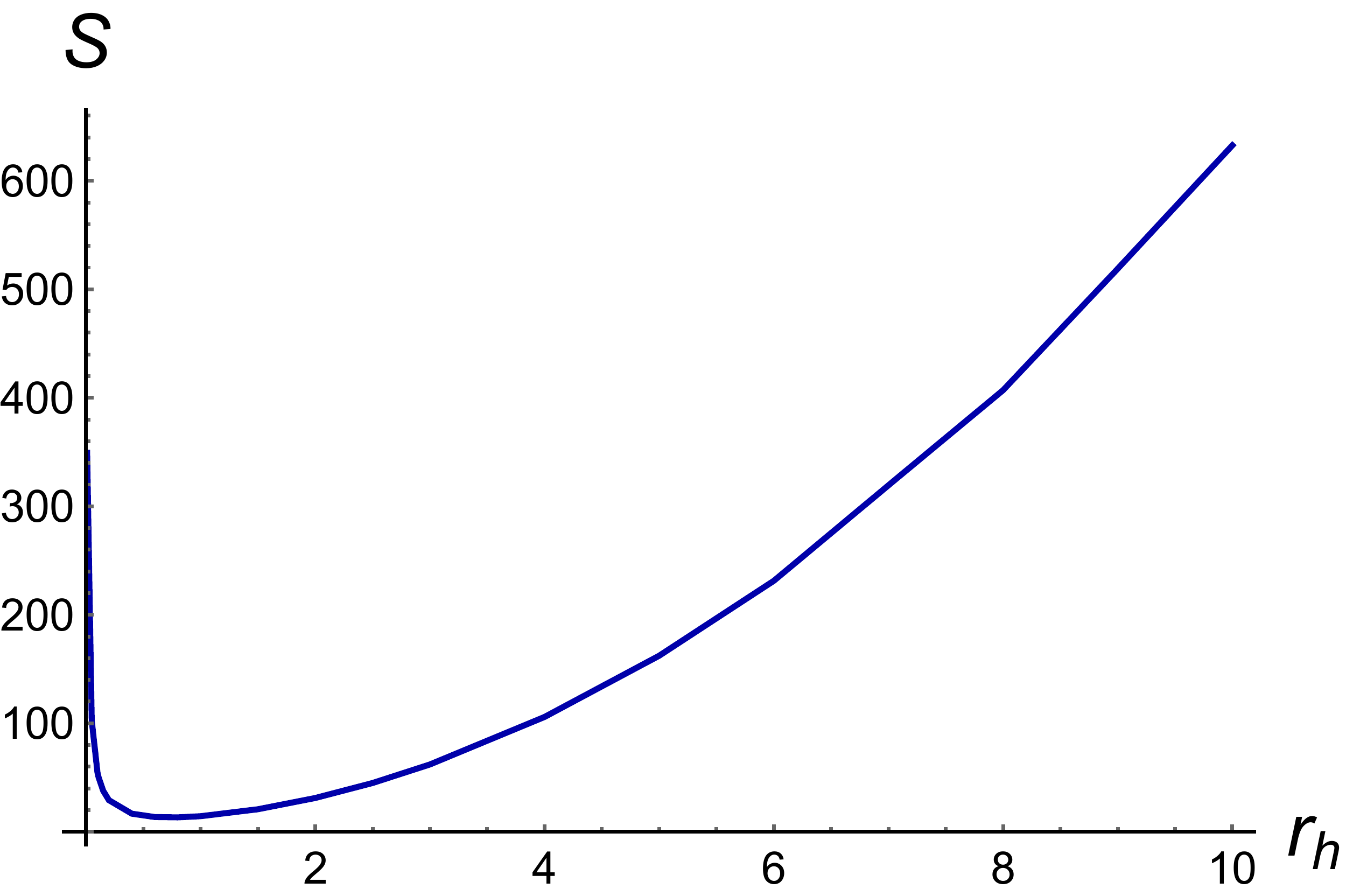}\qquad
\includegraphics[width=70mm]{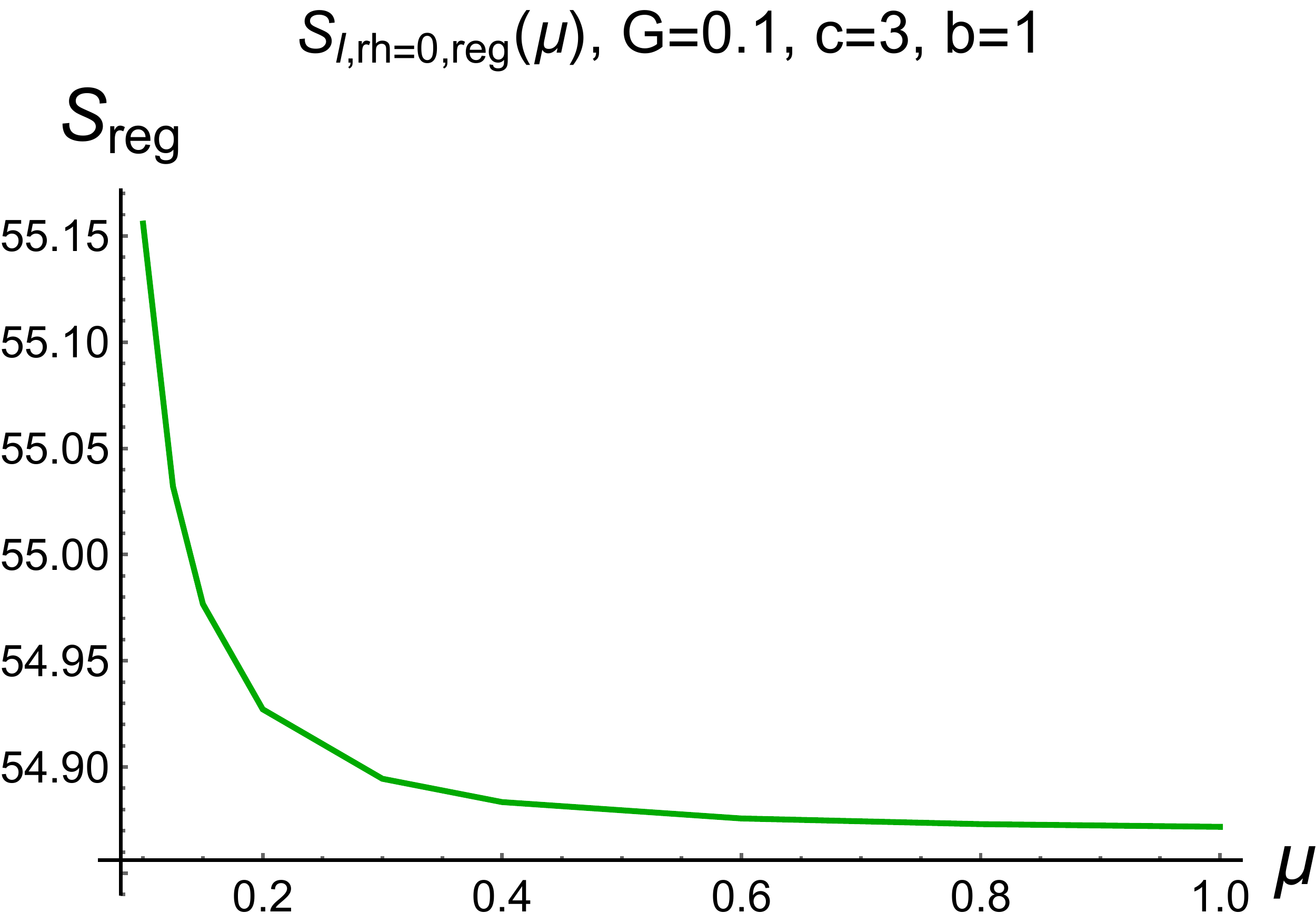}\caption{A) Dependence of  the regularized entropy  
$S_{Island}^{(B)}, B=2r_h+\mu$ on $M$,  here $\mu=0.01$.
%{\bf Math.file: solution-exact-8.nb, solution-exact-8 -Large-rh.nb}} 
B) Dependence of  the zero mass 
regularized   entropy $S_{Island}^{(B)}\Big|_{B=\mu}$ on regularization parameter $\mu$.}
%	{\bf Math.file: M00.nb.}
		\label{fig:Sregmu}
\end{figure}

   Extremizing \eqref{Sapr} on $a$ we get 
   %[[calculations in the file {\bf M00.nb}, previous calculations are in {\bf M0.nb}]]
\be
\frac{c \sinh \left(\frac{a-b}{G \mu
   }\right)}{3  \mu  \left(\cosh
   \left(\frac{a-b}{ \mu
   }\right)-1\right)}+4 \pi  a=0.\ee
Finding solution of this equation numerically we get the dependence of $ S_{\cI,reg,r_h=0}$
on regularization parameter $\mu$, Fig.\ref{fig:Sregmu}
We see, as should be,  this expression is singular for $\mu\to 0$.

\newpage

\section{Conclusion and discussion}
We  have considered the evaporation of the Schwarzschild black hole and note that, generally speaking,  the  island 
configurations do not provide a bounded entanglement entropy in the end of the black hole evaporation.  
 Despite the fact that
including 
the island and the extremization about its location results in  saturation of  the entropy for the eternal black hole,  the evaporating of black hole ends up by unbounded increasing of the entropy.  \\

One can argue that  the formula \eqref{island2} applies for a black hole held in equilibrium with radiation, and not for a freely evaporating black hole. The  second term $cb /6r_h$ can be interpreted as the entropy of infalling radiation\footnote{We thank Henry Maxfield for providing  this interpretation}.\\

 General arguments of possible  inconsistency of islands in theories with long-range gravity have been presented in 
\cite{Geng:2021hlu}.
\\

In \cite{Arefeva:2021byb} it was suggested
that the origin for such singular behavior lies in the use of the Kruskal coordinates.  The Kruskal coordinates
are suitable to the  analytical continuation of the  Schwarzschild metric but they are singular in  the limit of vanishing of the black hole mass $M\to 0$. Thus, the Kruskal coordinates are not appropriate to describe small black holes. 
   Depending of the initial state parameters the unboundeness  of the entanglement entropy can occur inside or outside of the Planck length.
 Similar remarks about islands and Page/anti-Page curves  are applied  to other static black holes, in particular, to Reissner-Nordstr\"om black holes \cite{Wang:2021woy, Kim:2021gzd}, black holes in  modified gravity \cite{Alishahiha:2020qza} as well as  black holes in dS space-time
\cite{Geng:2021wcq}.
\\

We have considered the regularization of applying the quantum extremal surface  prescription to gravity theory with matter fields to the  Schwarzschild black hole metric in $r_h\to 0$ using thermal  coordinates \cite{Arefeva:2021byb}. This consideration  permits to consider the entanglement entropy of island configurations  in the end of evaporation of the black hole, i.e. take the limit $r_h\to 0$. This regularization can be applied also to other static black holes. It would be also interesting to consider the modification of the Page curve in BCFT models of black hole \cite{Rozali:2019day,Sully:2020pza,Geng:2021iyq,Ageev:2021ipd} and holographic  moving mirror  in the end of evaporation \cite{Chen:2017lum,Akal:2020twv}. 
 \\

The consideration of small black hole presented here can be generalized to the Reissner-Nordstr\"om black holes \cite{Wang:2021woy, Kim:2021gzd},  charged dilaton black holes  in non-asymptotically flat cases \cite{Ahn:2021chg,Yu:2021cgi}
as well as to black holes in (A)dS  \cite{Geng:2021wcq}.  The latter are of particular interest in the context of possible phenomenological applications in the condense matter physics  \cite{Schalm:book,Erdmenger:book} and quark gluon plasma \cite{Arefeva:2014kyw}. In particular, the entropy for the dilaton charged black hole contains the $\log b/r_h$ term\footnote{We thank Hyun-Sik Jeong for taking our attention to this fact.}.
\\

%Note that the consideration in \cite{Hashimoto:2020cas}  and presented above is  rely on the assumptions  that  %Hawking radiation has no
%gravitational interaction and  the main contribution to the matter von Neumann entropy 
%comes from the entanglement between s-wave modes of the quantum field and one can reduce the theory into two 
% two dimensional conformal field theory. 

As a general remark, let us note that in \cite{Nieuwenhuizen:2005zq}  the black hole information problem has been considered as a particular
example of the fundamental irreversibility problem in statistical
physics. It was  pointed out that similar problem occurs when we study
ordinary gas or the formation of the ordinary black  body and
its thermal radiation. Actually, one has to give a quantum
mechanical explanation for the emergence of the second law of thermodynamics
in macroscopic systems. In this context one can say that  the Boltzmann equation for 
the one-particle distribution
function in the kinetic theory of gases is an analog of the extrimization equation for the entanglement entropy 
with island.\\

It can be assumed that if we take into account more detailed information about the dynamic properties of matter in the radiation region (not limited to the $s$-mode only) and use a more structured model of  islands, perhaps we will be able to more accurately  reconstruct the fine grained entropy of quantum gravity and avoid the problem with entropy growth in the end of the black hole evaporation. 

\section*{Acknowledgements} We would like to thank Dmitry Ageev, Michael Khramtsov,
Kristina Rannu,  Timofey Rusalev and Pavel Slepov 
for useful discussions.  We would also like  to thank Hyun-Sik Jeong, Nori Iizuka, Henry Maxfield and Suvrat Raju  for comments and correspondences.

\end{document}